\documentclass[twocolumn,           
               showpacs,            
               nopreprintnumbers,     
               aps,                 
               prd,          	    
               letterpaper,             
               groupeaddress,      
               nofootinbib,         
               tightenlines,        
               floats,floatfix      
               ]{revtex4-1}

\usepackage{graphicx}
\usepackage{dcolumn,soul}
\usepackage{bm}
\usepackage{amsmath}
\usepackage{amsfonts,amssymb}
\usepackage{color}
\usepackage{enumerate}
\usepackage{subfigure}
\usepackage{float}
\begin{document}

\title{Asymptotically Equivalent Functions and Ultrafilters applied to Noncommutative Quantum Cosmology}

\author{J. A. Astorga-Moreno$^{1}$}
\email{jesus.astorga@alumnos.udg.mx}

\author{E. A. Mena-Barboza$^{1}$}
\email{emena@cuci.udg.mx}

\author{Miguel A. Garc\'{\i}a-Aspeitia$^{2,3}$}
\email{aspeitia@fisica.uaz.edu.mx}

\affiliation{$^1$Centro Universitario de la Ci\'enega, Universidad de Guadalajara\\
Ave. Universidad 1115, M\'odulo de Investigaci\'on y Tutor\'ias, C.P. 47820 Ocotl\'an, Jalisco, M\'exico.}

\affiliation{$^2$Unidad Acad\'emica de F\'isica, Universidad Aut\'onoma de Zacatecas, Calzada Solidaridad esquina con Paseo a la Bufa S/N C.P. 98060, Zacatecas, M\'exico.}
\affiliation{$^3$Consejo Nacional de Ciencia y Tecnolog\'ia, Av. Insurgentes Sur 1582. Colonia Cr\'edito Constructor, Del. Benito Ju\'arez C.P. 03940, Ciudad de M\'exico, M\'exico.}

\date{\today}
\begin{abstract}

Using the semi-classical approximation to the Wheeler-DeWitt equation obtained via Arnowitt-Deser-Misner (ADM) formalism in the Friedmann-Lema\^itre-Robertson-Walker (FLRW) model coupled to a scalar field and positive cosmological constant, and in the Kantowski-Sachs (KS) Universe,  we introduced a deformation on the commutation relation for the minisuperspace variables and find an explicit semiclassical expression equivalent, in an adequate limit, to the solution with the aid of asymptotically equal functions and the theory of \emph{Ultrafilters}, offering a suggestive alternative to sketch the behavior of the dynamical system involved without the need to solve it numerically. 

 \end{abstract}
 \maketitle


\section{Introduction} \label{intro}

Perhaps the concordance or $\Lambda$-Cold Dark Matter ($\Lambda$CDM) model is the most successful via to reproduce the dynamics of our Universe \cite{Ade:2015xua} introducing the existence of dark entities known as dark matter (DM) and dark energy (DE). Despite its important achievements, the model suffers from different pathologies such as the flatness, the horizon problem (see \cite{Lyth:1998xn} for an excellent discussion), the current Universe acceleration \cite{Schmidt,Perlmutter,Riess}, among others. In particular, near or in the initial singularity and for times close to Planckian lengths and energies, it is necessary a quantum gravity theory (QG) to understand the universe in these regimes; in particular because gravity already plays a preponderant role at this quantum level. In this sense, the candidates to study this energy scale are for example the string theory \cite{Witten:1995ex} or loop quantum gravity \cite{Rovelli:2011eq}, being until now its predictions far from being falsifiable and its theoretical building far to be completely solved. These are some of the reasons of why the Wheeler-DeWitt (WDW) equation is still the cornerstone to address problems in QG scenarios and being also a great arena to pave the way in our search to find a final quantum gravity theory\footnote{It is important to notice that the WDW equation and its consequences are a limit case of the string theory.}.

The noncommutativity in coordinates (NCC) were introduced in the late of 40's  \cite{snyder}, generating a great deal of interest in this area of research \cite{connes1,connes,connes2,douglas,seib}, finding this \emph{boom} in the study of the effects in the phase-space of the classical and quantum cosmology (QC) \cite{vak,ncgrav}. It is noticeable to mention that the second case is a simplified approach to study the very early universe, where one could assume the effects of noncommutativity.
As it is well known, in general, the configuration space in QC (superspace) is infinite-dimensional, but for the homogenous cosmologies (like our Universe), where the metric depends only on time, we can obtain a model with a finite degrees of freedom called the \emph{minisuperspace}. In this context, like in noncommutative quantum mechanics \cite{bastos0,neto0,gam}, it is possible to introduce noncommutativity in a $2n$-dimensional configuration space via the change of variables (Boop-shift) which are often referred to as the Seiberg-Witten map \cite{seib}, and satisfy an extended Heisenberg algebra (see Appendix \ref{ap.b}). Therefore the traditional way to extract useful dynamical information and deal with the difficulties associated with solving the WDW equation in the challenging scenario of noncommutativity gravity is through the WKB-type method, and when it is used a $(2+1)$-minisuperspace model the wave function proposed takes the form shown in Appendix \ref{ap.b}; finding the associated Einstein-Hamilton-Jacobi (EHJ) equation, that is, a coupled system of two non-linear ordinary differential equations. To construct the EHJ equation and find analytical expressions, we observed the behavior, in a convenient limit, of functions involved addressing the analysis with a functions called asymptotic equals, (see Appendix \ref{ap.a}), selecting the appropiate candidate of the respective equivalence class.

As a demonstration for the functionality of the method, we will apply in the dynamical system derived from EHJ equation that comes from the FLRW metric, considering the noncommutativity in the space coordinates. In addition, together with the mentioned method we associate in the Noncommutative coordinate and momentum (NCCM)-KS Cosmology a particular family of subsets, called Ultrafilter, that is relevant in some branches of mathematics, like topology where in many cases are used to construct examples and counter examples \cite{gin,ber}, functional analysis and dynamical systems when discrete systems are studied \cite{sal,sal1}.  The study of cosmology in some limits (asymptotic analysis), is well known and appears in problems related with the cosmological constant \cite{wald, we}, the behavior near and far from an initial singularity in certain kinds of cosmologies \cite{mc, bill} and in quintessence models \cite{urena}.
Hence, the goal is support our work with the mathematical concepts mentioned and propose an analytical solution, unlike traditional methods studied in literature where is represented only in a integral form or solved numerically \cite{1eri,bastos}. Here we will take a step forward which will undoubtedly be useful for sketching the solution of the dynamical system associated to the EHJ equation from this approach.

This paper is organized as follows: In Sec. \ref{FLRW} we study the asymptotic behavior of functions in a semiclassical expression for the FLRW model with curvature $k \neq 0$ and cosmological constant ($\Lambda \neq 0$) in a NCC-frame. In Sec. \ref{KS} we obtain a semiclassical approach for the KS universe exploring a phase-space noncommutative extension and using again asymptotic analysis together with some properties of the collection of subsets introduced in the Appendix \ref{ap.a}, summarizing this procedure in Appendix \ref{ap.c}. By inspection, we compare our analytical expressions with the numerical solution of the dynamical system obtained from applying the semiclassical limit and our analysis in both models; also for the case of KS cosmology, with the numerical solution of the classical dynamical system. Finally, Sec. \ref{CO} deals with the conclusions and outlooks.
We will henceforth use units in which $c=\hbar=1$.

\section{Noncommutative FLRW Model} \label{FLRW}

In order to use the asymptotic behavior, we first proceed to study the EHJ equation that comes from FLRW cosmology with the presence of the parameter $\theta$. It is worth to notice that we will use FLRW model only as a laboratory to probe the effectiveness of the mathematical tool of asymptotically equal functions.  

We start this study using the line element in this background as:
\begin{equation}
ds^2=-\mathcal{N}(t)^2dt^2+a(t)^{2}\Big[\frac{dr^2}{1-kr^2} +r^2(d\vartheta^2+\sin^2(\vartheta) d\varphi^2)\Big], \label{frw}
\end{equation}
where $a(t)\equiv e^{ \alpha (t)}$ as the scale factor, $\mathcal{N}(t)$ is the lapse function and $k$ is the curvature constant.  

Based in previous work, Ref. \cite{1eri}, we calculate the canonical momenta for $\alpha$ and $\phi$ as 
\begin{equation}
\frac{d S_{2}}{d \alpha}=p_{\alpha}=12\frac{d  \alpha}{dt},  \qquad  \frac{d S_{1}}{d \phi}=p_{\phi}=-\frac{d  \phi}{dt}, \label{mo}
\end{equation}
and the classical Hamiltonian for the case $\Lambda\ne0$ and $k\ne0$ reads
\begin{align} 
H=\mathcal{N}\mathcal{H}=\mathcal{N}e^{-3\alpha}\Big[\frac{1}{24} p_{\alpha}^2 - \frac{1}{2}p_{\phi}^2+ e^{6\alpha}(2\Lambda+6ke^{-2\alpha})\Big],
\end{align}
the canonical quantization in the momenta and the Hamiltonian constraint for the ADM formulation, give us the WDW equation coupled to a scalar field and $\Lambda$ in the form:
\begin{equation} 
\Big[-\frac{1}{24} \frac{ \partial^2}{\partial \alpha^2} + \frac{1}{2}\frac{ \partial^2}{\partial \phi^2}+ e^{6\alpha}(2\Lambda+6ke^{-2\alpha})\Big]\psi(\alpha,\phi)=0,  \label{cuan}
\end{equation}
where $\mathcal{N}=e^{3\alpha}$ is chosen in order to fit the gauge. Since the effects of the deformation will reflect only in the WDW potential \cite{star}, when the noncommutativity in coordinates ($x^1=\phi$, $x^2=\alpha$) is applied in Eq. \eqref{cuan} and also the WKB-type method (Appendix \ref{ap.b}), we finally obtain the EHJ equation 
\begin{eqnarray}
\frac{1}{12} \Big[\Big(\frac{d S_{2}}{d \alpha}\Big)^2 +48\Lambda e^{6\alpha} +144ke^{4\alpha}+
(108\Lambda\theta e^{6\alpha}  +\nonumber\\  
216ke^{4\alpha} \theta)^2  \Big]  =
 \Big[\frac{d S_{1}}{d \phi}-\frac{(12\Lambda\theta e^{6\alpha}  +24ke^{4\alpha} \theta)}{2}\Big]^2,     \label{a1}
\end{eqnarray}
Hence, we deduce the equations
\begin{subequations}
\begin{eqnarray}
\frac{d S_{1}}{d \phi}&=&P_{\phi_0} \label{s1},\\
\frac{1}{12}\Big(\frac{d S_{2}}{d \alpha}\Big)^2&=&P^2_{\phi_0} - 4\Lambda e^{6\alpha}C^a_{\theta} -12ke^{4\alpha}C^b_{\theta}.\label{s2}
\end{eqnarray}
\end{subequations}
with $P_{\phi_0}$ a positive decoupling constant and $C^a_{\theta}, C^b_{\theta}$ also constants (depending on $\theta$) satisfying 
\begin{equation}
\lim_{ \theta \to 0}C^\mu_{\theta}=1, \quad \lim_{ \theta\to +\infty}C^\mu_{\theta}=+\infty.
\end{equation}
Therefore applying \eqref{mo}  in \eqref{s1}-\eqref{s2} and choosing $C^a_{\theta}=e^{ 3\theta P_{\phi_0}}$, $C^b_{\theta}=e^{ 2\theta P_{\phi_0}}$, the system can be written in the form
\begin{subequations}
\begin{eqnarray}
d \phi&=&-P_{\phi_0}dt,\label{intb} \\
t-t_0&=&\sqrt{12} \int \frac{d\alpha}{[ P_{\phi_0}^2  - e^{6\alpha}(4\Lambda e^{ 3\theta P_{\phi_0}} +12ke^{-2\alpha}e^{  2\theta P_{\phi_0} })]^{1/2}}.\label{int}
\end{eqnarray}
\end{subequations}
The initial condition $\phi(t_0)=\phi_0$ give us $\phi(t)=\phi_0-P_{\phi_0}(t-t_0)$. For Eq. \eqref{int} defined in the interval $(-\infty,\alpha_0)$ with $\alpha_0<0$, under inspection, we have for the integrand
\begin{equation}
G_{ \theta }( \alpha)=\Big[  P_{\phi_0}^2  - e^{6\alpha}(4\Lambda e^{ 3\theta P_{\phi_0}} +12ke^{  2\theta P_{\phi_0} }e^{-2\alpha})\Big]^{-1/2},
\end{equation}
the following asymptotically equal function
\begin{equation}
F_{ \theta }( \alpha)=\Big(  P_{\phi_0}^2  - A^2_{\theta}e^{6\alpha} \Big)^{-1/2},
\end{equation}
with $A_{\theta}=[ 4\Lambda e^{ 3\theta P_{\phi_0}} +12ke^{  2\theta P_{\phi_0} } ]^{1/2}$. Hence using \eqref{asymp1} and \eqref{asymp}, considering decreasing $\alpha$'s, the expression is: 
\begin{equation}
\alpha=  \frac {1}{3 }\ln\Big[ \frac {P_{\phi_0}}{A_{\theta} }\rm{sech}\Big(\frac {\sqrt{3}A_{\theta}P_{\phi_0}(t-t_0)}{2}\Big)\Big],  \label{alpha} 
\end{equation}
where it is applied the initial condition $6\alpha(t_0)= \ln\Big( P^{2}_{\phi_0}/A^{2}_{\theta} \Big)$. As a complement, the other variable has the form:
\begin{equation}
\phi=\phi_0-P_{\phi_0}(t-t_0)-\sqrt{3}\theta P_{\phi_0}  \rm{tanh}\Big(\frac{\sqrt{3}P_{\phi_0}( t-t_0 )}{2} \Big).\label{phi0} 
\end{equation}
Proceeding in the same manner we obtain the commutative expressions for the general form as
\begin{subequations}
\begin{eqnarray}
&\phi(t)=\phi_0-P_{\phi_0}(t-t_0),  \label{c}\\
&\alpha(t)= \frac {1}{3 }\ln\Big[ \frac {P_{\phi_0}}{A_{0} }\rm{sech}\Big(\frac {\sqrt{3}A_{0}P_{\phi_0}(t-t_0)}{2}\Big)\Big], \label{c1}\\
&A_{0}=\sqrt{ 4\Lambda +12k    }.
\end{eqnarray}
\end{subequations}

\begin{figure*}[h]
\centering
\subfigure[Commutative]{\includegraphics[width=0.40\textwidth]{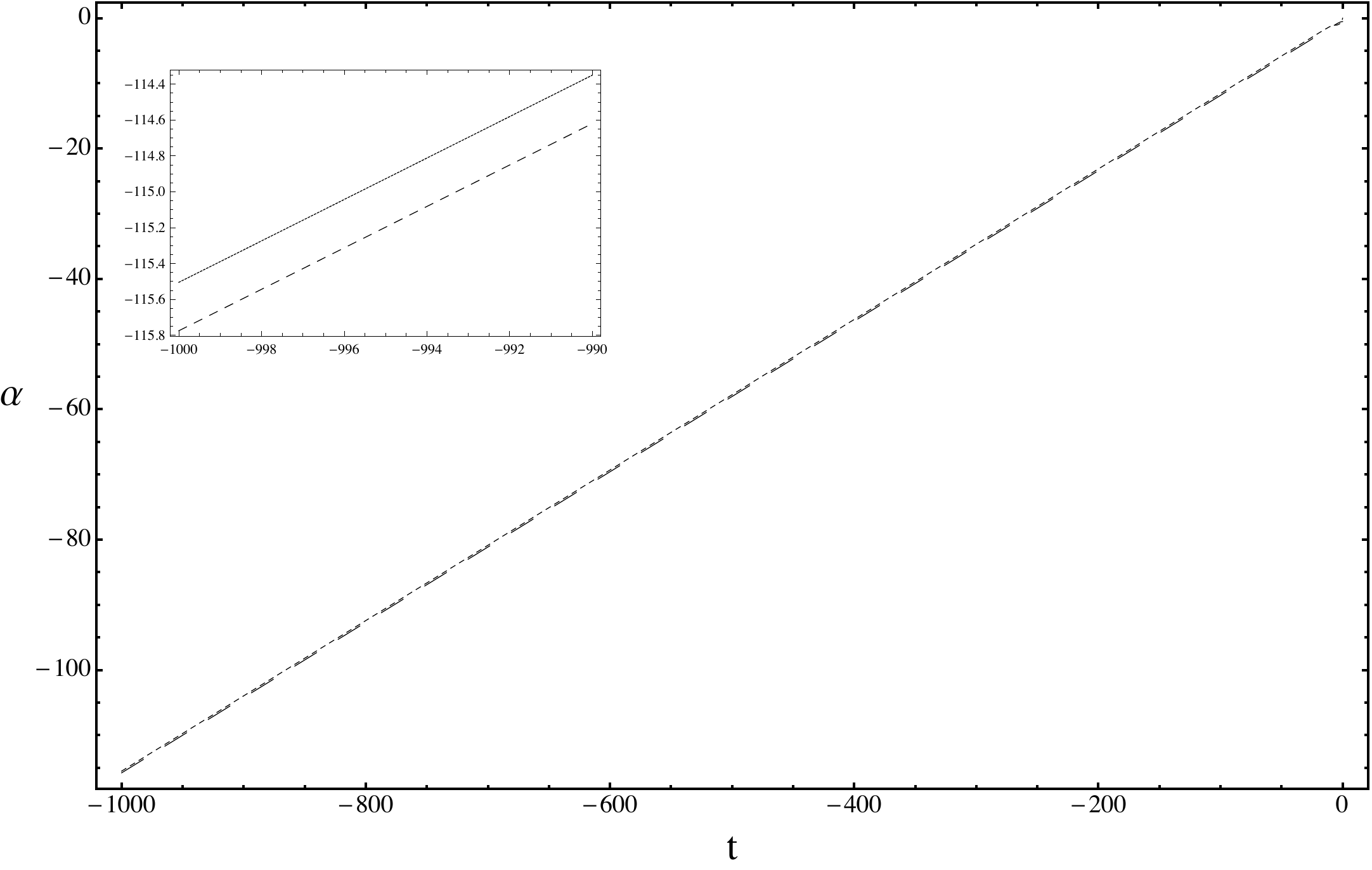}} \label{om01}  
\subfigure[NCC]{\includegraphics[width=0.40\textwidth]{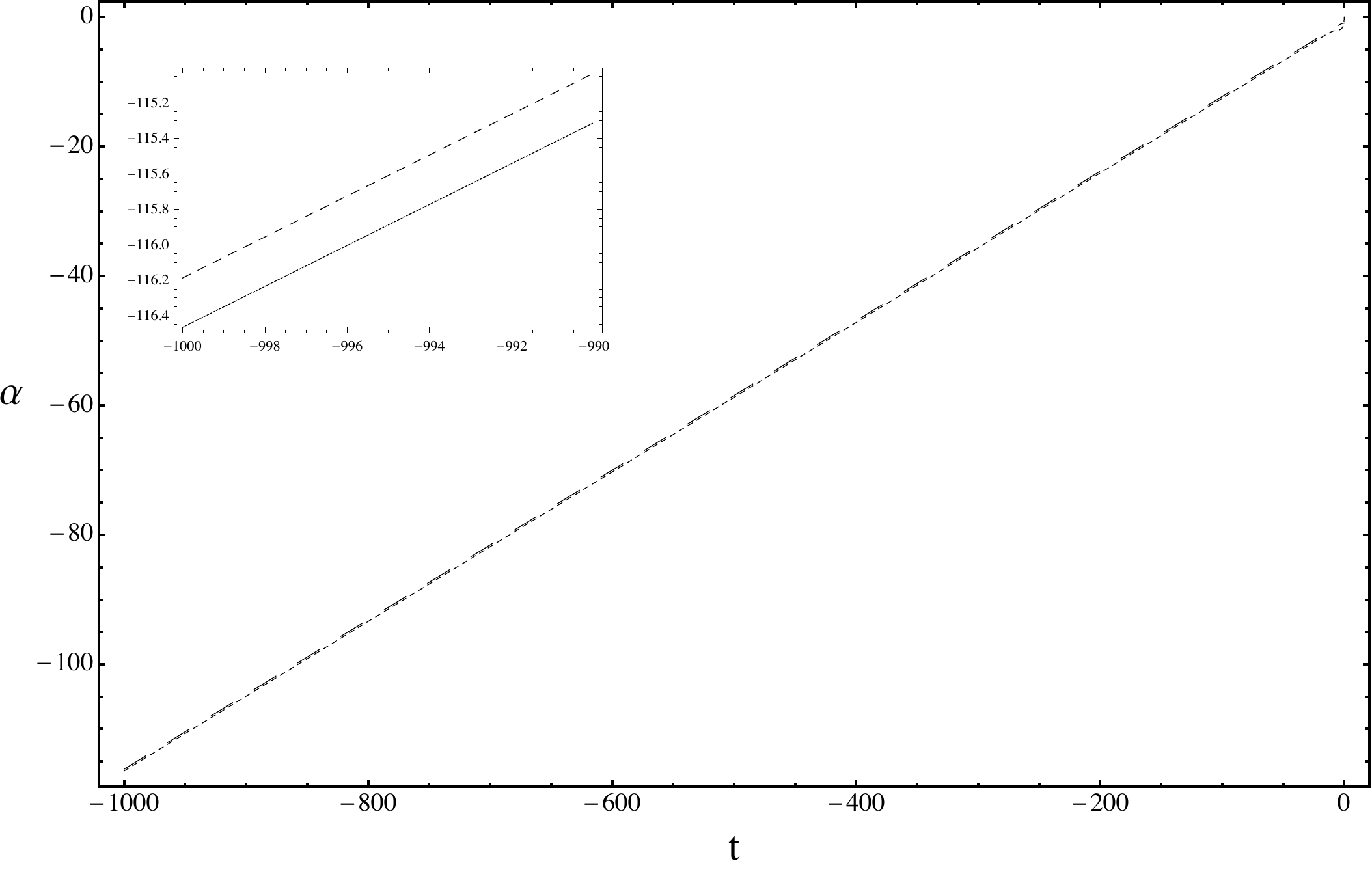}} \label{om02}  
\caption{Plots for $\alpha$, comparing in (a) the analytical expression \eqref{c1} (dashed line) and numerical solution of Eqs. \eqref{intb} and \eqref{int} with $\theta=0$ (pointed line) and in (b) the analytical expression obtained in Eq. \eqref{alpha} (dashed line) with the solution of Eqs. \eqref{intb} and \eqref{int} in a NCC frame ($\theta=5$), both under $t_0=0$, $P_{\phi_{0}}= 2/5$, $ \Lambda=k=1$ and the initial conditions $\alpha(t_0)=-2.30259$ and $\alpha(t_0)=-4.77981$, respectively.} \label{om1}
\end{figure*}

\begin{figure*}[h]
\centering
\subfigure[Commutative]{\includegraphics[width=0.40\textwidth]{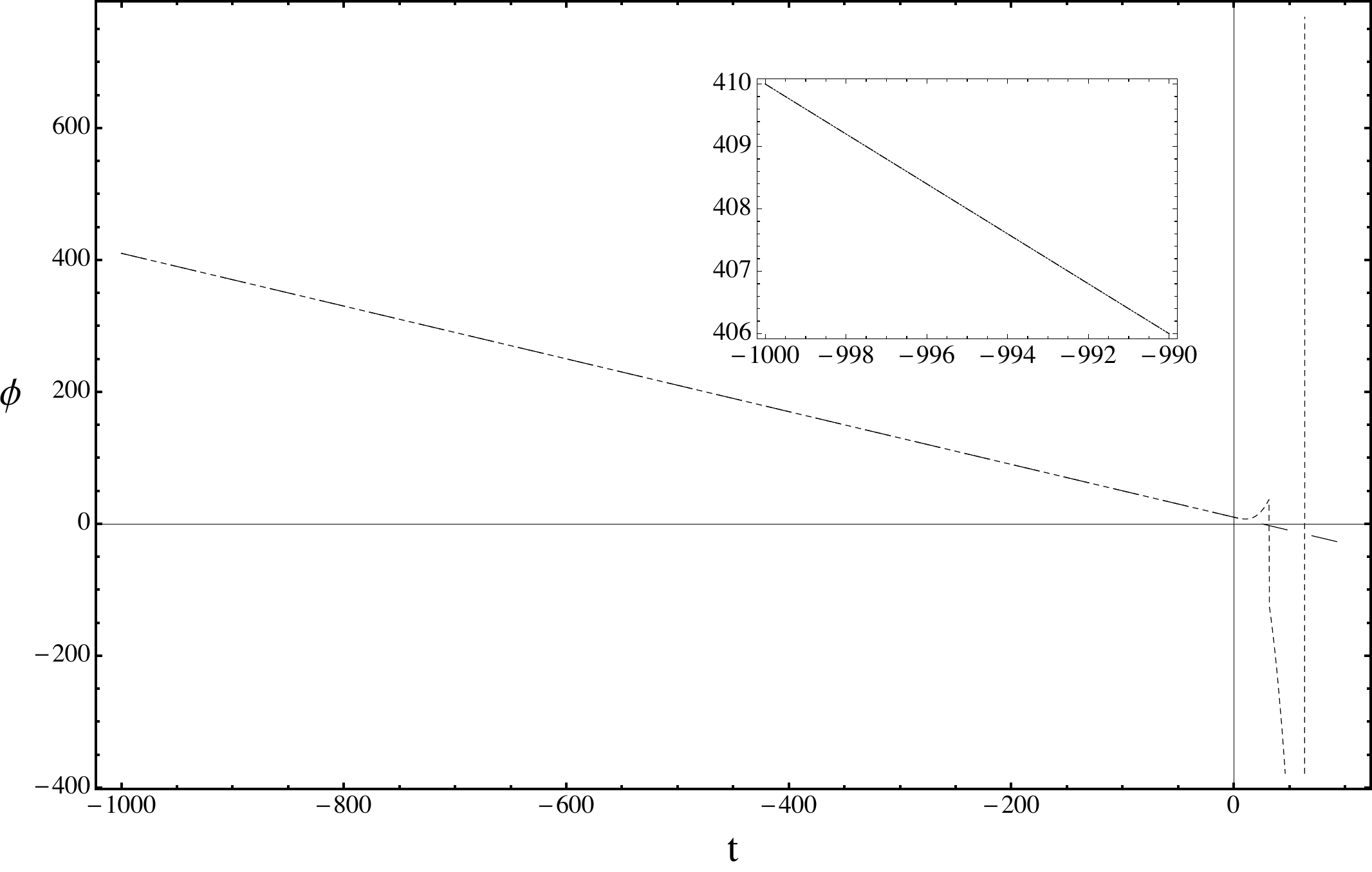}} \label{phi01}
\subfigure[NCC]{\includegraphics[width=0.40\textwidth]{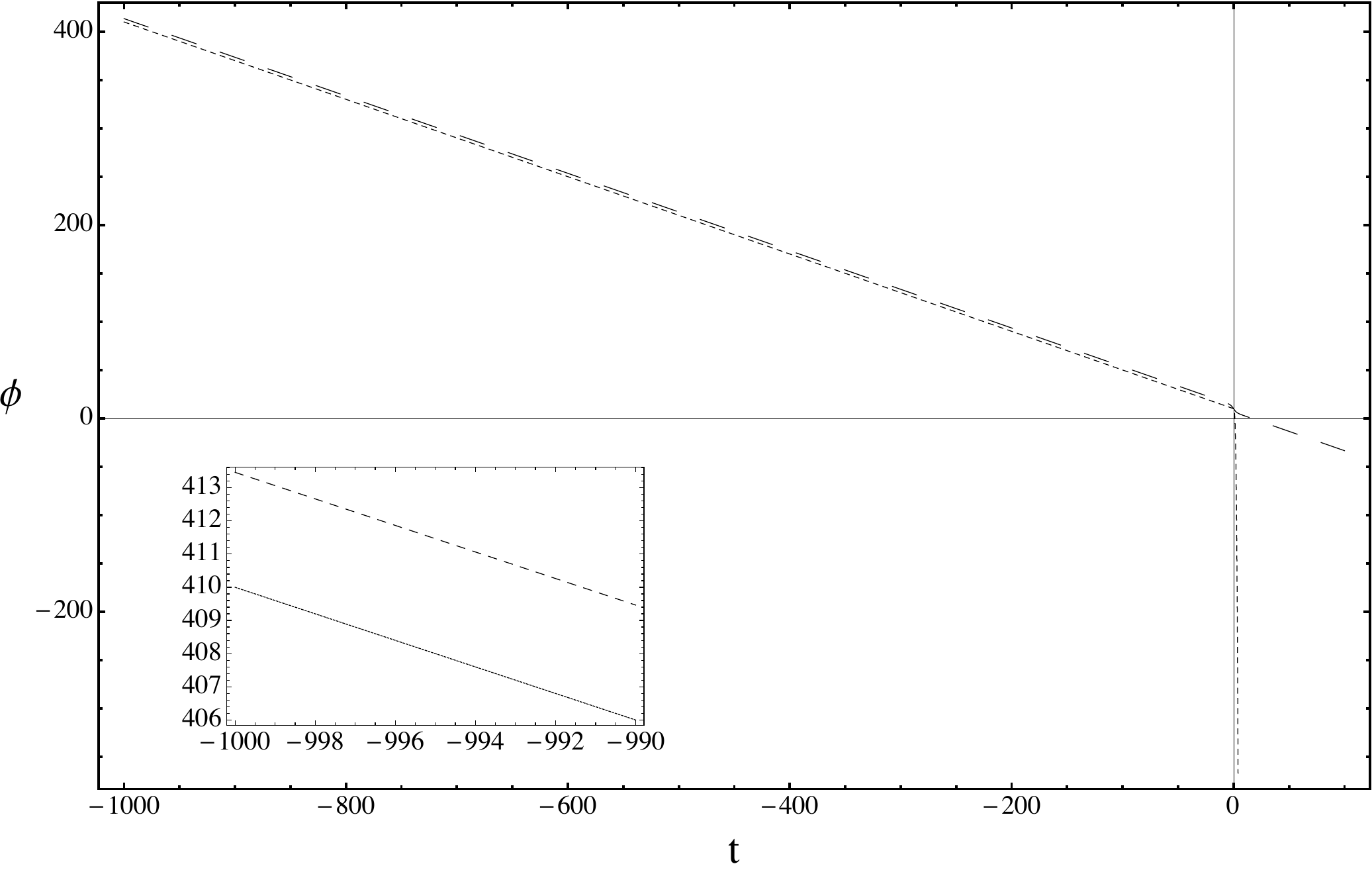}} \label{phi02}
\caption{Plots for $\phi$, comparing in (a) the analytical expression \eqref{c} (dashed line) and numerical solution of Eqs. \eqref{intb} and \eqref{int} with $\theta=0$ (pointed line) and in (b) the analytical expression obtained in Eq. \eqref{phi0} (dashed line) with the solution of Eqs. \eqref{intb} and \eqref{int} in a NCC scenario ($\theta=5$) both under the same conditions that Fig. \ref{om1}. The initial condition is $\phi(t_0)=\phi_0=10$.} \label{phi}
\end{figure*}

\begin{figure*}[h]
\centering
\subfigure[Commutative]{\includegraphics[width=0.40\textwidth]{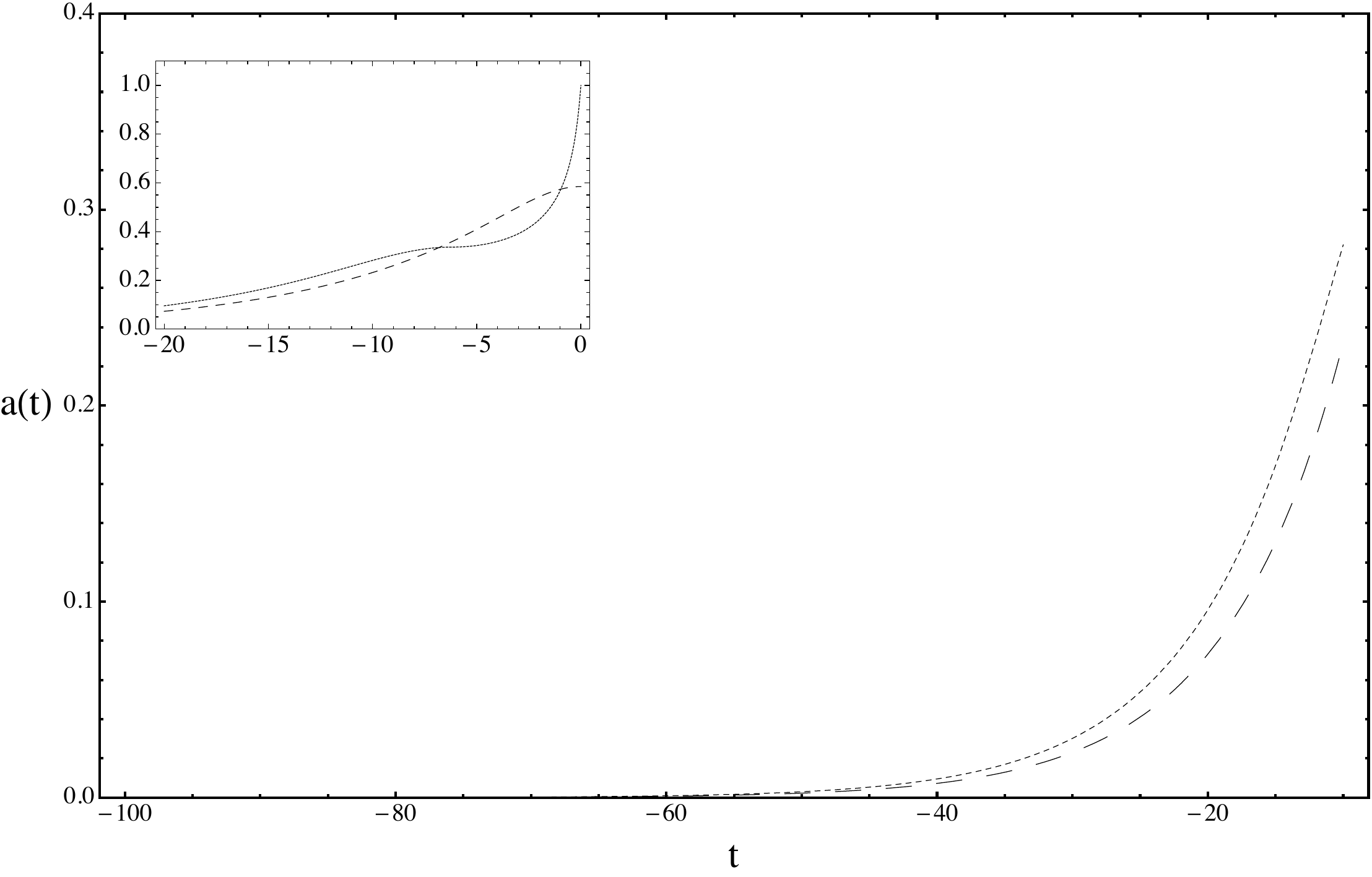}} \label{om201}
\subfigure[NCC]{\includegraphics[width=0.40\textwidth]{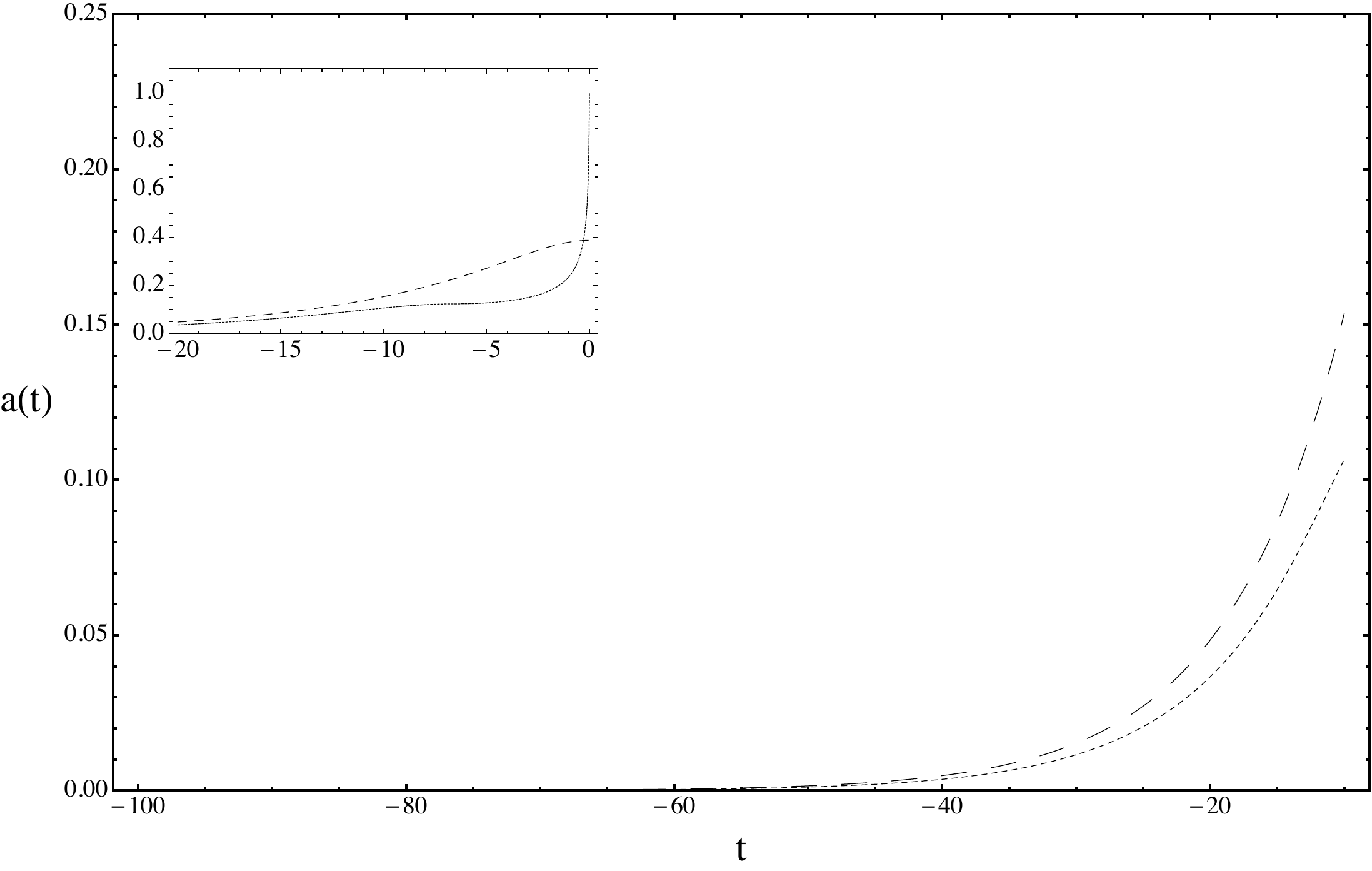}} \label{om202}
\caption{Plots of  the scale factor $\exp( \alpha ) $ in a commutative (a) and a NCC (b) frame using the numerical solution of Eqs. \eqref{intb}, \eqref{int} (pointed line) and the expressions \eqref{intb}, \eqref{alpha} (dashed line). For the commutative frame we make $\theta=0$ in the system and the same conditions that in Fig. \ref{om1} are considered.} \label{om2}
\end{figure*}

In Table \ref{tab} we present, for decreasing values of $t$, the relative error (that give the error rate (ER) between the numerical solution of the system \eqref{intb}-\eqref{int} (with $\theta=0$ and $\theta\neq0$) and the respective analytical expressions \eqref{c} and \eqref{c1} for the commutative frame and \eqref{alpha} and \eqref{phi0} for the NCC one. Then Fig. \ref{om1} and Fig. \ref{phi} shows the plots for values of $t$ in the interval $[-1000,0]$ observing that correspond to $\alpha$'s in $[-120,0]$ and $\phi$'s in $[0,400]$. In Fig. \ref{om2} the factor that measures the evolution\footnote{This is the scale factor $a(t)=e^{\alpha}$.} is analyzed, plotting this parameter with the analytical $\alpha$ and the numerical one in a commutative and NCC scenario, showing that when $t$ is decreasing they are all similar. For $a(t)$, when the values of $t$ are near zero, the plots in both frames lies in the range $[0,1]$ (See the  internal boxes in Fig. \ref{om2}).
\begin{table}[htbp]
\begin{center}
\begin{tabular}{|c|c|c|c|c|}
\hline
$t$ & ER $\alpha_{conm}$'s& ER $\phi_{conm}$'s& ER $\alpha_{ncc}$'s&ER $\phi_{ncc}$'s \\
\hline \hline
-100&$ 2.33029\% $&$ 9.94\times 10^{-14}\% $&$ 2.22202\% $&$ 6.9282\% $ \\ \hline
-700& $ 0.33\%$ &$ 7.84\times 10^{-14}\%$&$ 0.340638\%$&$ 1.19452\%$ \\ \hline
-800&$ 0.29207\% $&$ 6.89\times 10^{-14}\%  $&$ 0.298513\% $& $ 1.04973\% $\\ \hline
-900&$ 0.259629\% $&$  6.14\times 10^{-14}\% $&$ 0.26566\% $&$ 0.936244\% $\\ \hline
-950&$ 0.245969\% $&$  5.83\times 10^{-14}\%  $&$ 0.251804\% $&$ 0.88823\% $\\ \hline
-990&$ 0.236034\% $&$  5.60\times 10^{-14}\%  $&$ 0.241718\% $&$ 0.853227\%$\\ \hline
-1000&$ 0.233674\% $&$  5.54\times 10^{-14}\%  $&$ 0.239321\% $&$0.844903\%$\\\hline
-2000&$ 0.11685\% $&$  2.80\times 10^{-14}\% $ &$ 0.120175\% $& $ 0.427667\%$\\ \hline
\end{tabular}
\caption{Error rate between $\alpha$'s, $\phi$'s commutative and $\alpha$'s, $\phi$'s NCC (numerical and analytical expressions) with $t_0=0$, $P_{\phi_{0}}= 2/5$, $\theta=5$, $ \Lambda=k=1$ and the initial conditions $\alpha(t_0)=-2.30259$ (commutative frame), $\alpha(t_0)=-4.77981$ (NCC frame) and $\phi(t_0)=\phi_0=10$. See the text for more details.}
\label{tab}
\end{center}
\end{table}

\section{Noncommutative KS Universe} \label{KS}

In this section we are not only going to apply an asymptotic treatment, also we will introduce the concepts\footnote{In this case we need an additional argument to pick the corresponding asymptotically equal function.} shown in Appendix \ref{ap.a} looking analytical expressions for the associated dynamical system that comes for the noncommutative space-space and momentum-momentum variables of the EHJ equation in the KS universe.

We start with the line element in the Misner parametrization \cite{mis}:
\begin{equation}
\begin{array}{cc}
 ds^2=-\mathcal{N}^2dt+X^2(t)dr^2 +Y^2(t)(d\vartheta^2+\sin^2(\vartheta) d\varphi^2),\\\\
X(t)=e^{-\sqrt{3}\beta},  \quad Y(t)=e^{-\sqrt{3}\beta}e^{-\sqrt{3}\Omega},
\end{array}
\end{equation}
where $X$ and $Y$ are the scale factors. Following the same recipe as in Sec. \ref{FLRW}, the WDW equation \cite{ncgrav} is:
\begin{equation}
\Big(\frac{\partial^2}{\partial \Omega^2}-\frac{\partial^2}{\partial \beta^2}-48e^{-2\sqrt{3}\Omega}\Big)\psi(\beta,\Omega)=0. \label{wdwks}
\end{equation}
As the authors of Ref. \cite{1eri} find, the momenta are
\begin{equation}
\frac{d S_{2}}{d \Omega}=p_{\Omega}=-\frac{1}{2}\frac{d  \Omega}{dt}, \quad  \frac{d S_{1}}{d \beta}=p_{\beta}=\frac{1}{2}\frac{d  \beta}{dt},  \label{momks}
\end{equation}
and remembering Appendix \ref{ap.b}, with the presence of the parameters $\theta$ and $\eta$ ($x^1=\beta$, $x^2=\Omega$), give the equation 
\begin{eqnarray}
      &&\Big(\frac{d S_{2}}{d \Omega} - \frac{\eta \beta}{2}   \Big)^2 +48e^{-2\sqrt{3}\Omega} +R(\Omega)\nonumber\\&=& \Big[\frac{d S_{1}}{d \beta}+\frac{ (\eta\Omega +48\sqrt{3} \theta e^{-2\sqrt{3}\Omega})}{2} \Big]^2, \label{ii}
\end{eqnarray}
where $R(\Omega)$ is defined as:
\begin{equation}
  R(\Omega)=96\sqrt{3} \sigma \Omega e^{-2\sqrt{3}\Omega}+ \frac{ (48\sqrt{3} \theta e^{-2\sqrt{3}\Omega})^2}{4},
\end{equation}
being $\sigma$ as in the first appendix.

\begin{figure*}[h]
\centering
\subfigure[$\Omega_A$  versus numerical solution of \eqref{sys0}, \eqref{sys}.]{\includegraphics[width=0.40\textwidth]{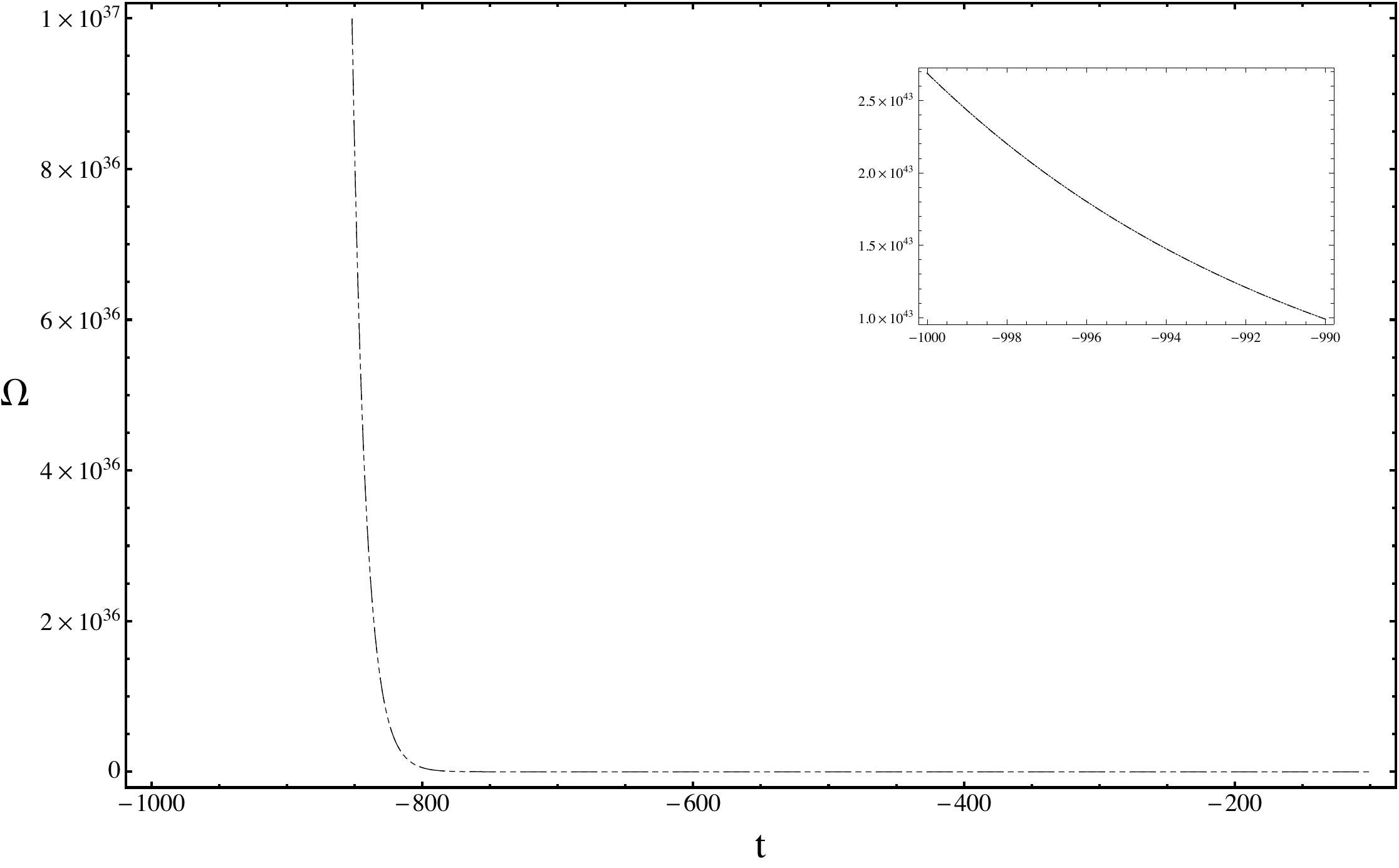}} \label{o101}
\subfigure[$\beta_A$ versus numerical solution of \eqref{sys0}, \eqref{sys}.]{\includegraphics[width=0.40\textwidth]{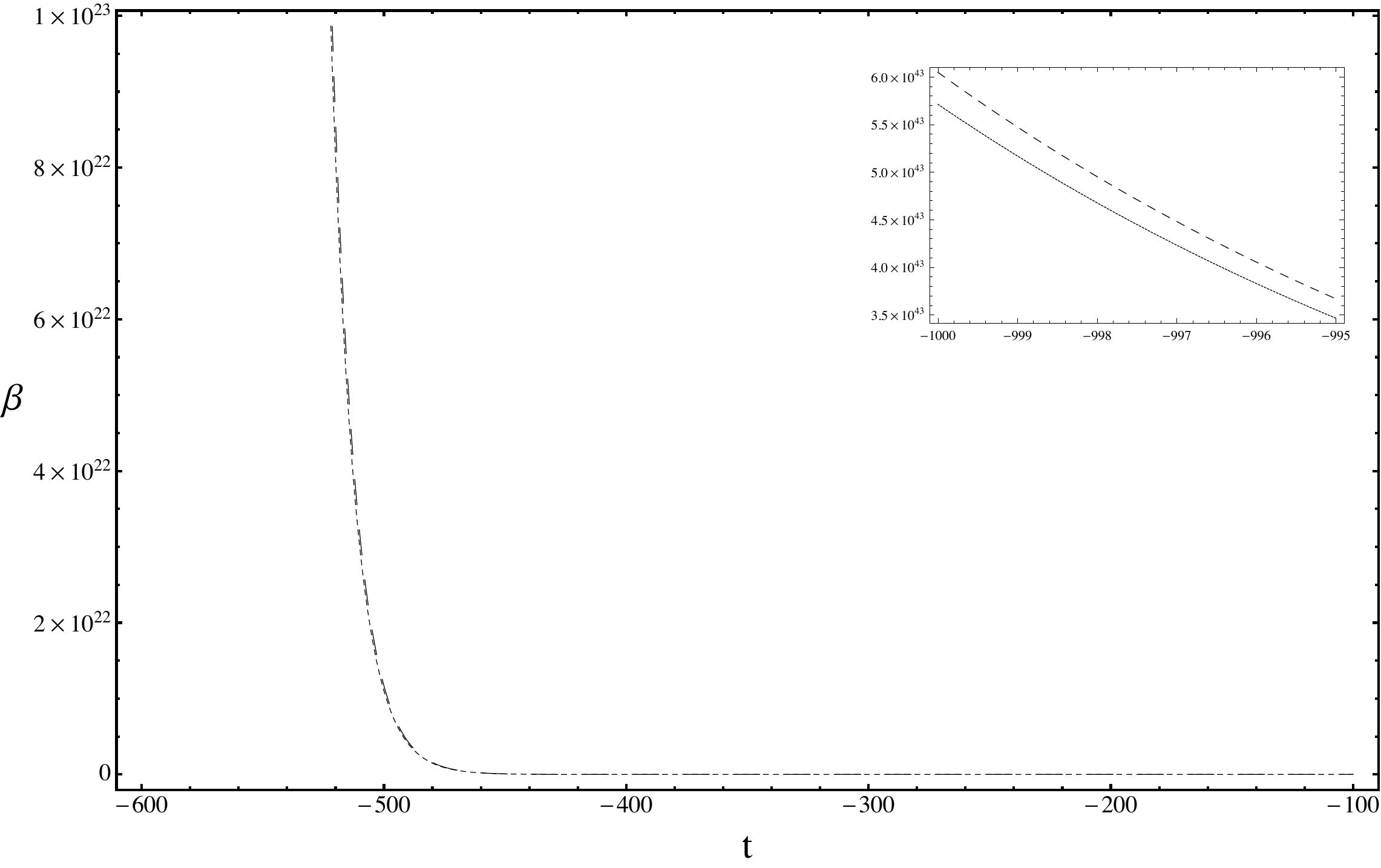}} \label{o102}
\caption{Plots of $\Omega_A$ and $\beta_A$ (dashed line) comparing with numerical solutions of \eqref{sys0}, \eqref{sys} (pointed line), where $t_0=-100$, $P_{\beta_{0}}= 2/5$, $\theta=5$, $\eta=0.1$, $C=1.5\times 10^{-5}$ also considering the initial conditions $\Omega(t_0)=\Omega_0=2.20265\times10^4$ and $\beta(t_0)=\beta_{0}=4.68142\times10^4$.} \label{o1}
\end{figure*}

In addition, noticing that $\exp(-2\sqrt{3}\Omega)=o(1) $\footnote{ The notation $o(1)$ represents: $\lim_{\Omega  \to +\infty}\exp(-2\sqrt{3}\Omega)/1=0$, then taking $\epsilon=1 \times 10^{-n}$ with $n\geq 2$ a natural number, we have that $\Omega>\frac{1}{2\sqrt{3}}\ln\Big( \frac{1}{1 \times 10^{-n}} \Big)$. For example, if $n=5$ we get $e^{-2\sqrt{3}\Omega}<1\times 10^{-5}$ when $\Omega>3.3235.$
}  when $\Omega  \to +\infty$, we consider that for large $\Omega$'s the first expression in Eq. \eqref{ii} can be written in the form
\begin{eqnarray}
    &&  \Big(\frac{d S_{2}}{d \Omega} - \frac{\eta \beta}{2}   \Big)^2 +48e^{-2\sqrt{3}\Omega}+R(\Omega) \nonumber\\&\approx &
      \Big( \sqrt{\Big(\frac{d S_{2}}{d \Omega}\Big)^2+48e^{-2\sqrt{3}\Omega}+R(\Omega)}- \frac{\eta \beta}{2}  \Big)^2,\label{aprox}
\end{eqnarray}
hence using the last approximation in \eqref{ii} can be derived the system of equations:
\begin{subequations}
\begin{eqnarray}
\frac{d S_{1}}{d \beta}&=&P_{\beta_0} \label{s11}-\frac{\eta\beta}{2}\label{s11}\\
\Big(\frac{d S_{2}}{d \Omega}\Big)^2&=&\Big(P_{\beta_0} +\frac{\eta\Omega}{2}\Big)^2- 48 e^{-2\sqrt{3}\Omega}E_{\theta}.\label{s22}
\end{eqnarray}
\end{subequations}
where $ P_{\beta_0}$ is like in Section II and $E_{\theta}$ is a constant with the property 
\begin{equation}
\left\lbrace
\begin{array}{ll}
E_{\theta}=1,\quad \theta \to 0\\\\
 E_{-\theta}\gg1, \quad \theta \to +\infty. 
\end{array}
\right.
\end{equation}
With the relations \eqref{momks} and taking $E_{\theta}=e^{-\sqrt{3}\theta P_{\beta_{0}} } $ we obtain the dynamical system
\begin{subequations}
\begin{eqnarray}
 d  \beta &=& (2P_{\beta_0}-\eta \beta)dt \label{sys0},\\
 d  \Omega&=&-2\Big[ \Big(P_{\beta_{0}}+\frac{ \eta \Omega}{ 2}  \Big)^2 \nonumber\\&&- 48e^{-2\sqrt{3}\Omega}e^{-\sqrt{3}\theta P_{\beta_{0}} } \Big]^{1/2}dt. \label{sys}
\end{eqnarray}
\end{subequations}
The solution of Eq. \eqref{sys0} with $\beta(t_0)=\beta_0$, give us
\begin{equation}
 t-t_0=  \ln \Big(   \frac{ 2 P_{\beta_{0}}-\eta\beta_{0} }{ 2 P_{\beta_{0}}-\eta\beta  }   \Big)^{1/\eta},   \label{ks1}
\end{equation}
and for the parameter $t$ it is possible to express in quadratures as
\begin{align}
t&=  -\frac{1}{2} \int \left[ \Big(P_{\beta_{0}}+\frac{ \eta \Omega}{ 2}  \Big)^2 - 48e^{-2\sqrt{3}\Omega}e^{-\sqrt{3}\theta P_{\beta_{0}}} \right]^{-1/2} d\Omega,   \nonumber\\
&= -\frac{1}{2} \int G_{\theta,\eta}d\Omega. \label{ks2}
\end{align}
In the case where $\eta \to 0$, equations \eqref{ks1} and \eqref{ks2} are reported in literature (see \cite{1eri} for details). To extract an analytical representation for the minisuperspace variables, in Appendix \ref{ap.c} we suggest a mathematical procedure to obtain an asymptotically equal function for $G_{\theta,\eta}$. Therefore, using \eqref{asymp} and \eqref{final}, we find the expression
\begin{equation}
\Omega_A(t)=e^{-\eta(t-C) }\label{e},
\end{equation}
considering $\Omega(t_0)=\Omega_0$ give $C= (\ln(\Omega_0)+\eta t_0)\eta^{-1}$ and
\begin{eqnarray}
\beta_A(t)=\frac{1}{\eta}\Big[2P_{\beta_0}-e^{-\eta(t-t_0) }\Big(2P_{\beta_0}-\eta\beta_0\Big)\Big]+\nonumber\\
\sigma e^{-\eta(t-C) }.\label{e1}
\end{eqnarray}
The behavior of the above expressions versus the numerical solution of the system  \eqref{sys0}-\eqref{sys} for $t \in [-1000,-100]$ (or values of $\Omega$ in the interval $ [22026.5,10^{37}]$) are in the Fig. \ref{o1}, where similarities for small $t$'s are notorious. For example, taking $t=-1000, -990, -950, -900. -800$, we have for $\Omega$ that the quotient between the numerical solution and the analytical expression is $1.00038$ corresponding to a relative error (or ER) of $0.037\%$. Making the same for $\beta$ we get that the quotients are $0.94444$ and the error rate produced is $5.8824\%$. From the above we can consider that the numerical solution and $\Omega_A$ are asymptotically equivalent when $t\to -\infty$, finding a similar behavior for $\beta_A$.

\begin{figure*}[h]
\centering
\subfigure[$q_{n,t}=1.80811$.]{\includegraphics[width=0.35\textwidth]{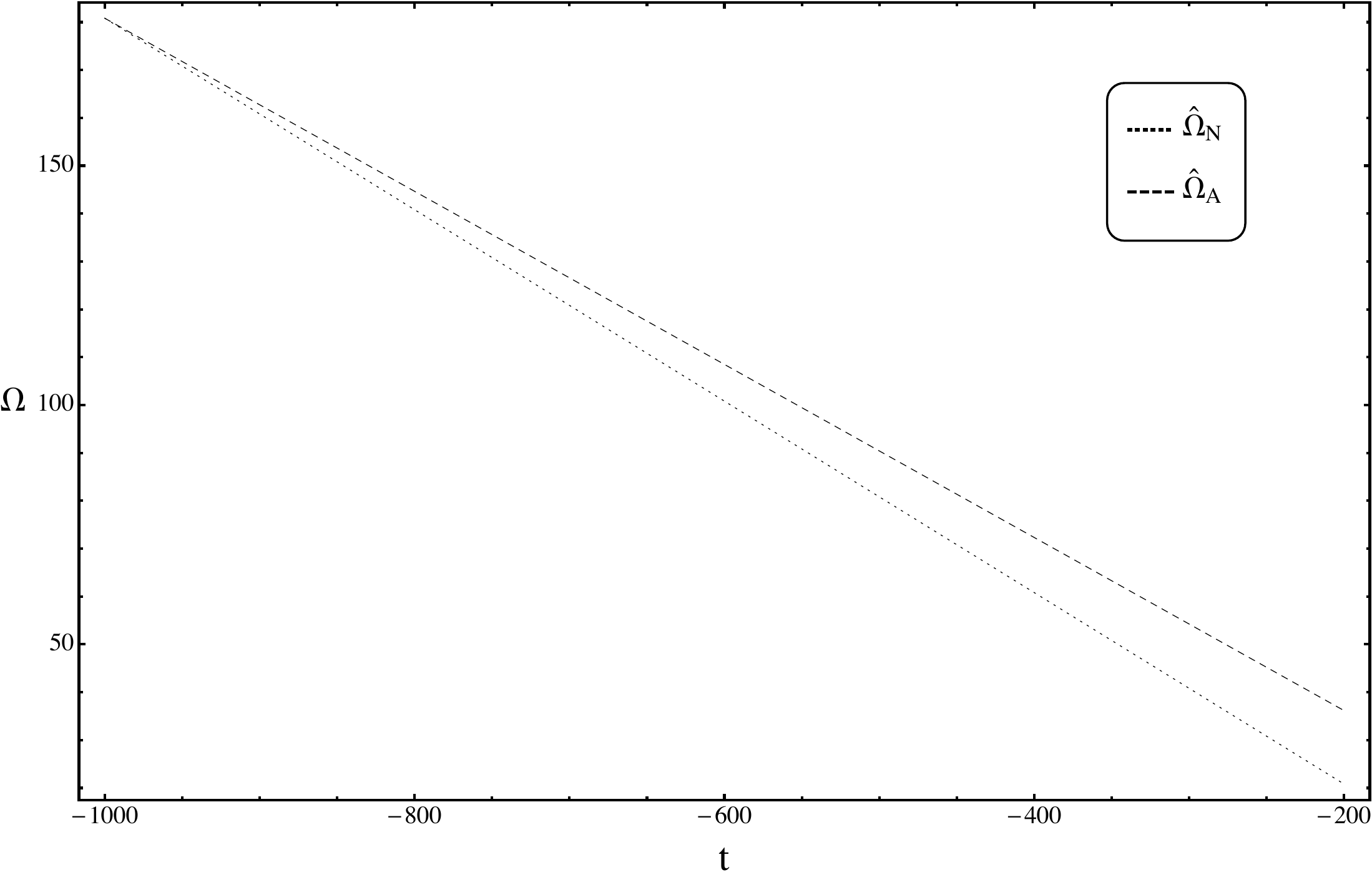}} \label{cla1}
\subfigure[$q_{n,t}=1.90406$.]{\includegraphics[width=0.35\textwidth]{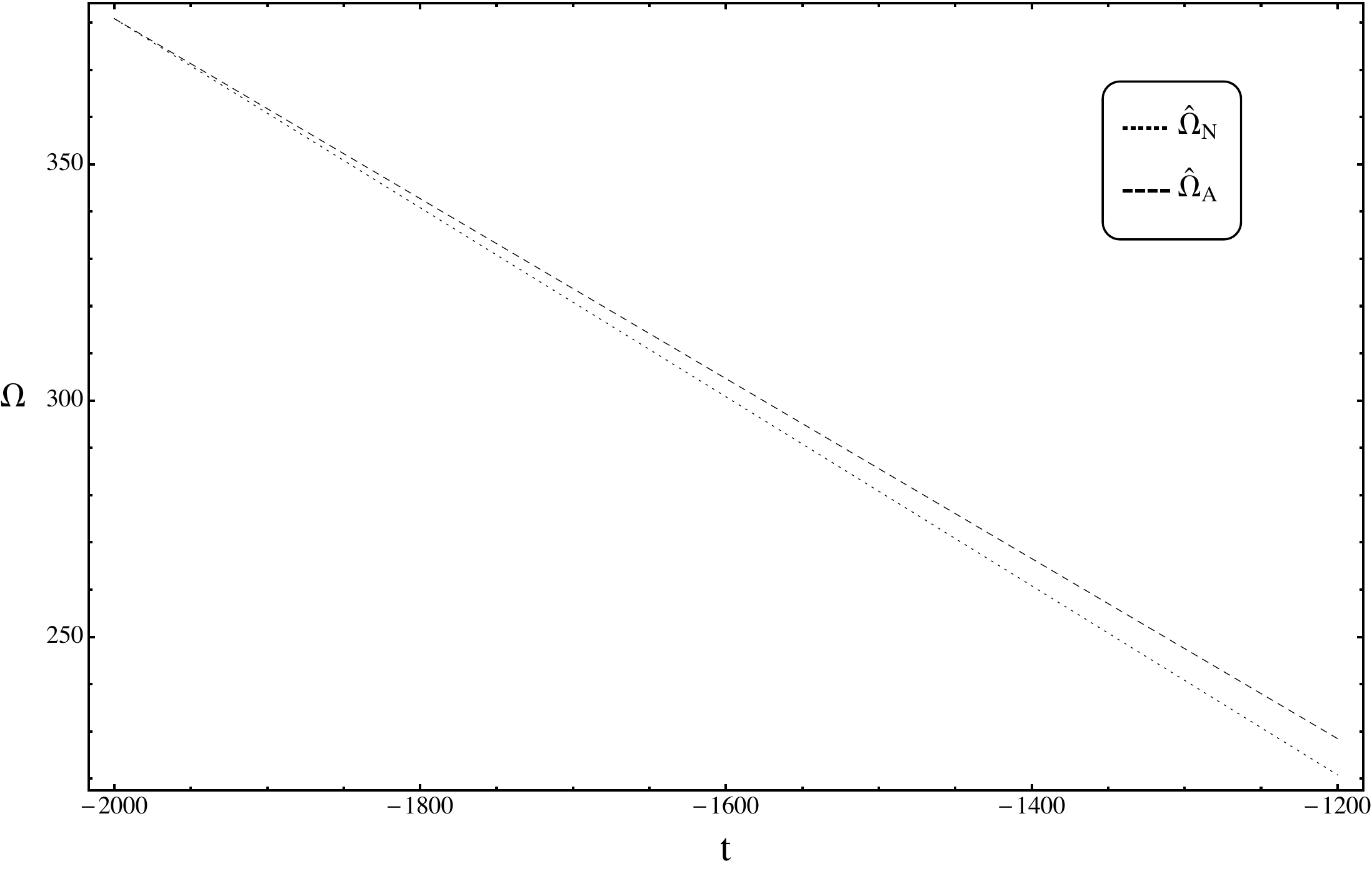}} \label{cla2}
\subfigure[$q_{n,t}=1.93604$.]{\includegraphics[width=0.35\textwidth]{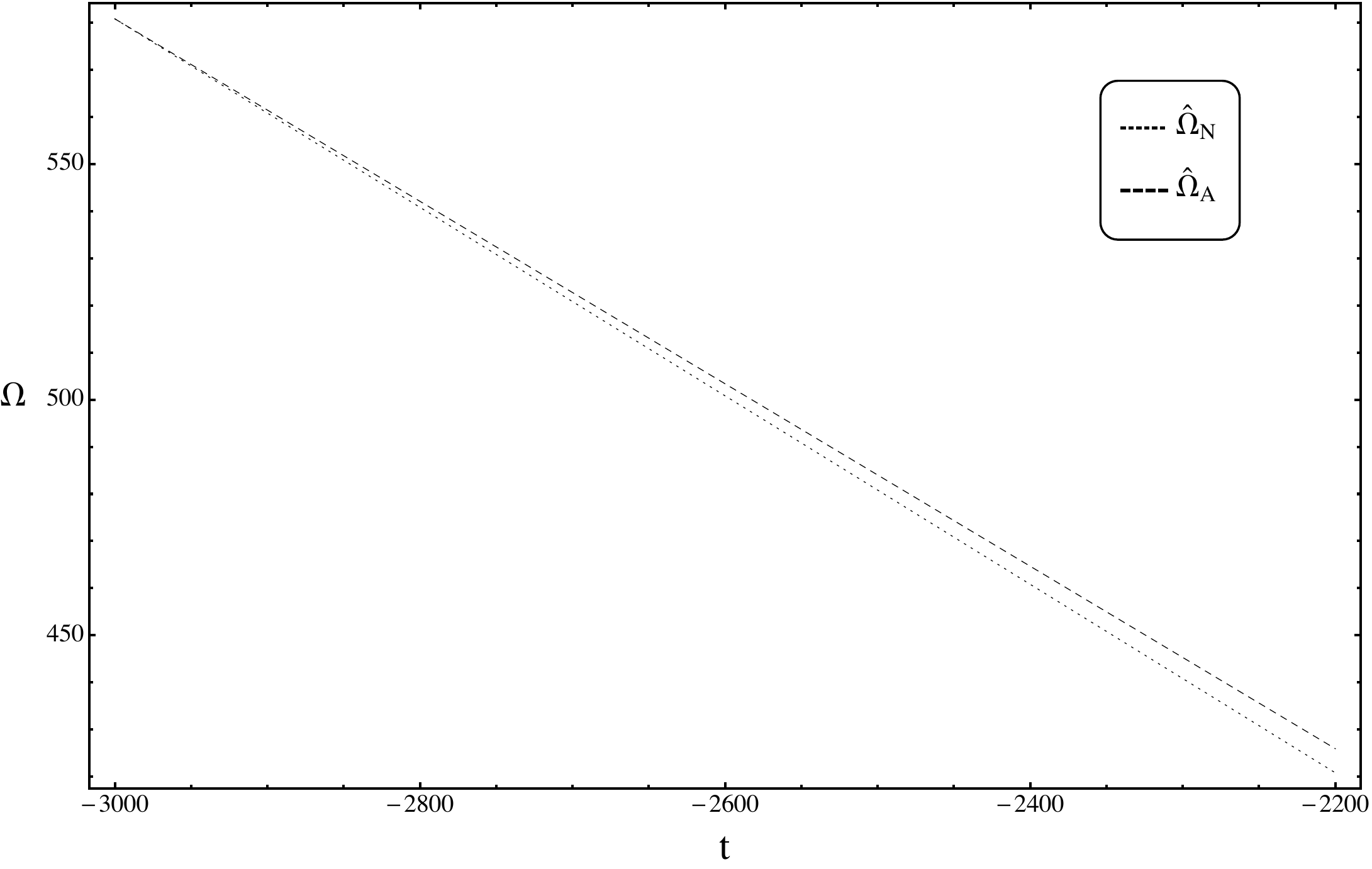}} \label{cla3}
\caption{Plots of $\hat{\Omega}_N$ (dotted line) and $\hat{\Omega}_A$ (dashed line) in the intervals $[-1000,-200]$ with $q_{n,t}=1.80811$, $[-2000,-1200]$ with $q_{n,t}=1.90406$ and $[-3000,-2200]$ with $q_{n,t}=1.93604$,
  where $t_0=-100$, $\theta=5$, $\eta=0.1$, $C=1.5\times 10^{-5}$ also considering the initial conditions $\Omega(t_0)=\Omega_0=2.20265\times10^4$ and $\beta(t_0)=\beta_{0}=4.68142\times10^4$, $P_{\Omega}(t_0)= 0$, $P_{\beta}(t_0)= 2/5$.} \label{cl1}
\end{figure*}

\begin{figure*}[h]
\centering
\subfigure[$q_{n,t}=1.80811$.]{\includegraphics[width=0.35\textwidth]{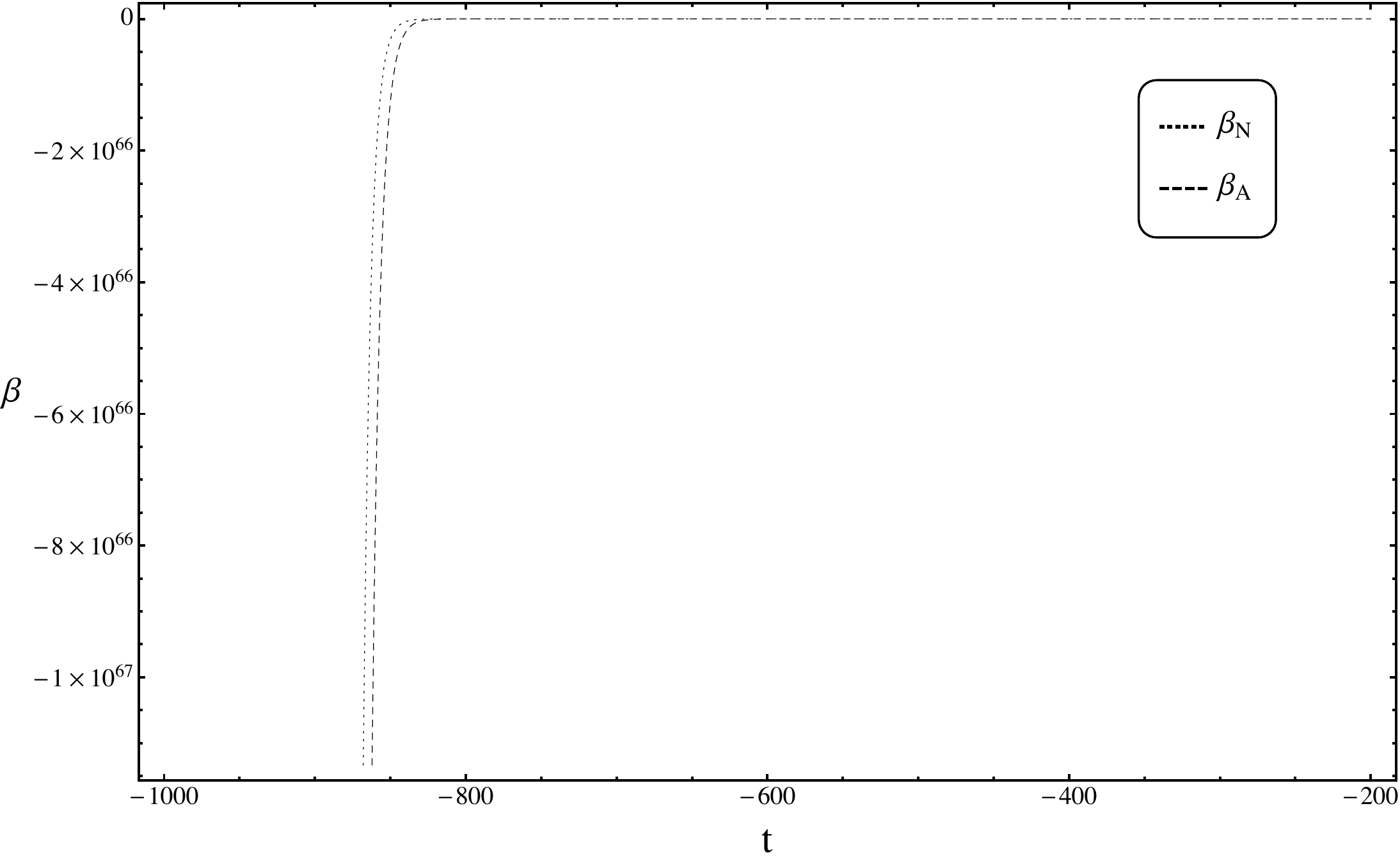}} \label{b1}
\subfigure[$q_{n,t}=1.90406$.]{\includegraphics[width=0.35\textwidth]{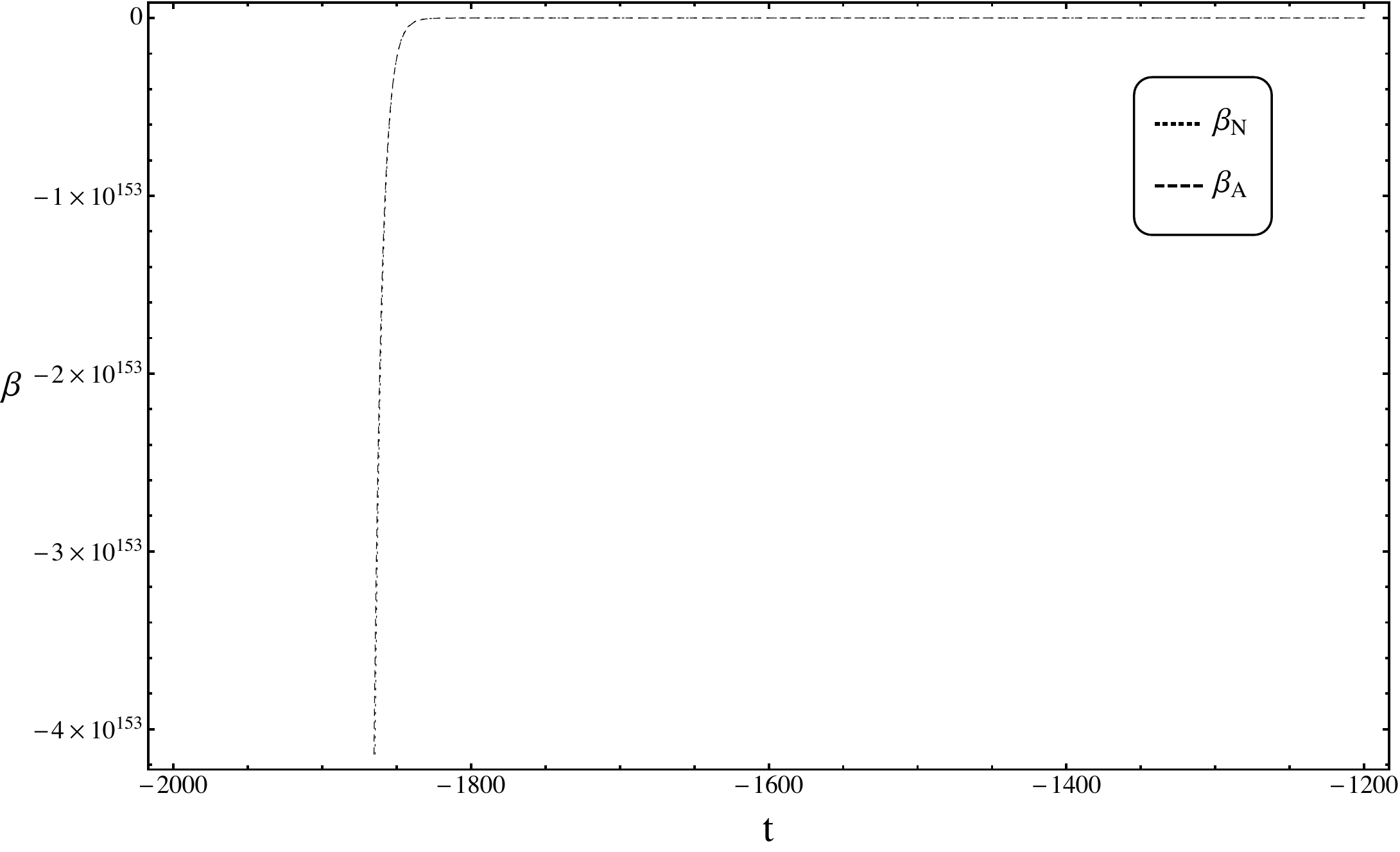}} \label{b2}
\subfigure[$q_{n,t}=1.93604$.]{\includegraphics[width=0.35\textwidth]{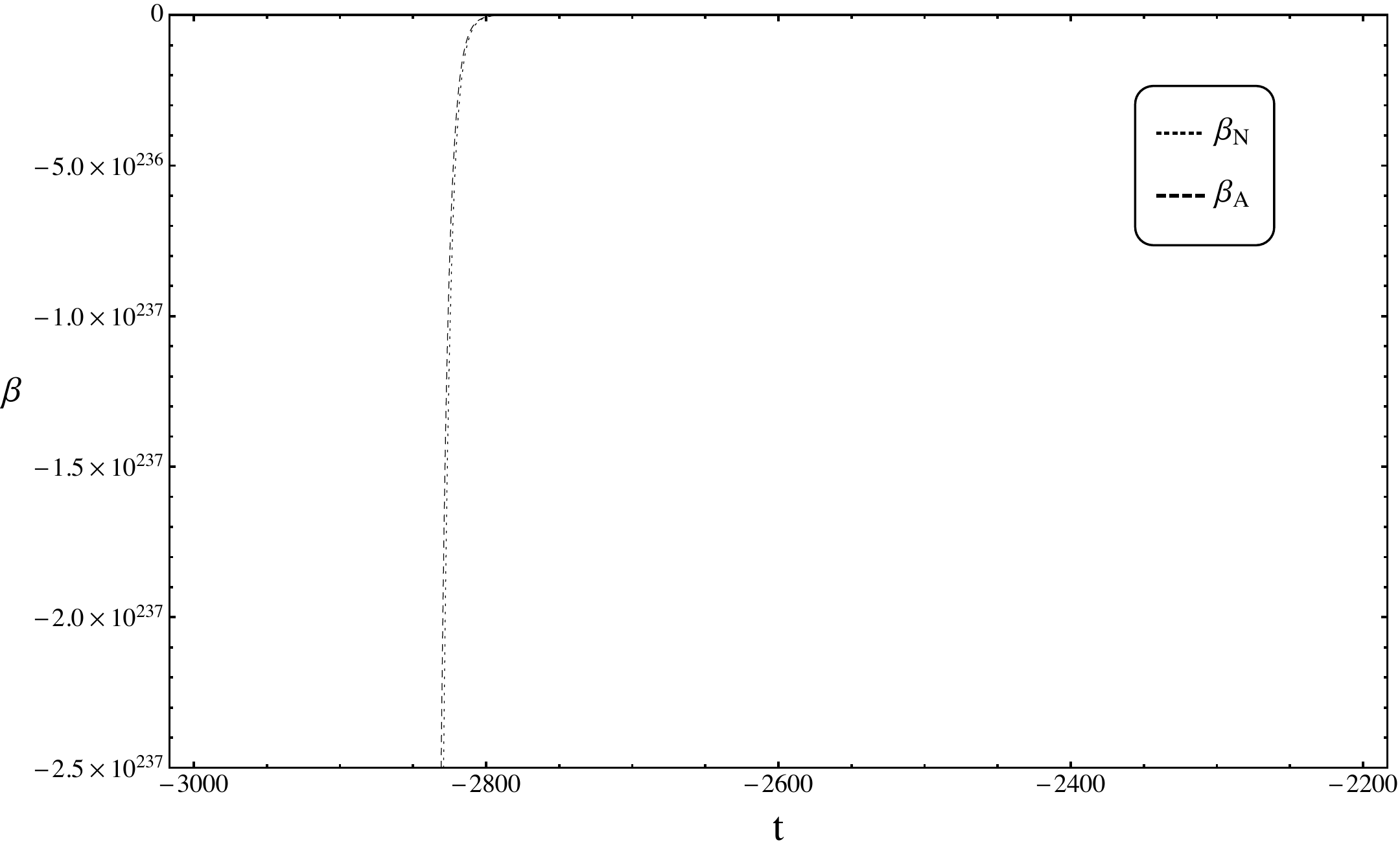}} \label{b3}
\caption{Plots of $\Omega_N$ (dotted line) and $\Omega_A$ (dashed line) in the intervals $[-1000,-200]$ with $q_{n,t}=1.80811$, $[-2000,-1200]$ with $q_{n,t}=1.90406$ and $[-3000,-2200]$ with $q_{n,t}=1.93604$,
  where $t_0=-100$, $\theta=5$, $\eta=0.1$, $C=1.5\times 10^{-5}$ also considering the initial conditions $\Omega(t_0)=\Omega_0=2.20265\times10^4$ and $\beta(t_0)=\beta_{0}=4.68142\times10^4$, $P_{\Omega}(t_0)= 0$, $P_{\beta}(t_0)= 2/5$. } \label{beta}
\end{figure*}

Now, the noncommutative relations imposing between the coordinates and their momenta in the modified Poisson algebra as
\begin{subequations}
\begin{eqnarray}
&\{\beta,\Omega\}=\theta, \quad \{p_\beta,p_\Omega\}=\eta,\label{p1}\\
&\{\Omega,p_\Omega\}=\{\beta,p_\beta\}=1+\sigma,\label{p21} 
\end{eqnarray}
\end{subequations}
giving the classical equations of motion:
\begin{subequations}
\begin{eqnarray}
&\dot{\Omega}=\{\Omega,H\}=-2(1+\sigma)p_{\Omega},\label{ccl1}\\
&\dot{\beta}=\{\beta,H\}=2(1+\sigma)p_{\beta}+96\sqrt{3}\theta e^{-2\sqrt{3}\Omega},\label{cl2} \\
&\dot{p}_{\Omega}=\{p_{\Omega},H\}=-(1+\sigma)96\sqrt{3}\theta e^{-2\sqrt{3}\Omega}-2\eta p_{\beta} , \label{cl3}\\
&\dot{p}_{\beta}=\{p_{\beta},H\}=-2\eta p_{\Omega}. \label{cl4} 
\end{eqnarray}
\end{subequations}
An analytical solution of this system is beyond reach given the distributions of the variables involved, hence, in Fig. \ref{cl2} and \ref{beta} we present the numerical solutions $\Omega_{N},\beta_{N}$ for Eqs. \eqref{ccl1}-\eqref{cl4}.
Let $\hat{\Omega}_N=\ln(\Omega_N)$ and $\hat{\Omega}_A=\ln(\Omega_A)$, we observe that the quotients of this functions, using the data in Fig. \ref{o1}, approximately satisfy
\begin{equation}
  q_{n,t}=\frac{\hat{\Omega}_{N}(t)}{\hat{\Omega}_A(t)}\approx 1+r_n, \quad n  \in \mathbb{N}
  \end{equation}
for $t\in I_n=[-200-25n,-200-25(n-1))$, where
  \begin{equation}
 r_n=\sum_{i\leq n, t_i\in I_i}(a_{i,t_i}\times 10^{-1}), \quad a_{i,t_i} \in \{0,1\},\label{r1}
 \end{equation}
 and $t_n=t$. The sequence $\{a_{i,t_i}\}$ is divergent but have a notorious property:  If $\{\rvert Z_i\lvert\}$ is the sequence of cardinalities of the sets $Z_i$ of consecutive zeros (or the significative number is the same), we note that it also diverges since the number of consecutive zeros increases as $t$ decreases\footnote{For the established values we have that in the interval  $[-4000,-200]$ the cardinalities of the sets  $Z_i$ are $\{1,2,3,5,12,37,83\}$. } and it is possible to treat $q_{n,t}$ as a constant when $t \to -\infty$\footnote{The difference between $q_{n,t_{1}}, q_{n,t_{2}}$ is equal or less than $10^{-1}$, when $t_{i} \in J_{i}$ and $J_{i}$ are consecutive intervals with $\rvert J_{i}\lvert=975$.}, then under these assumptions we have
 \begin{equation}
  \hat{\Omega}_{N}(t)  \approx q_{n,t}\hat{\Omega}_A(t) \quad \text{or} \quad \Omega_N(t)\approx e^{-\eta q_{n,t}(t-C) }.
\end{equation}
In Table  \ref{tab1} we check the error rate (ER) between $\hat{\Omega}_N$ and $\hat{\Omega}_A$ for values in $-6000\leq t \leq -200$, also for every $q_{n,t}$ we consider that $t$ is the minimum value in the interval $I_n$ and observing that as $t$ decreases so does the relative error\footnote{ Here we consider a refined value of $q_{n,t}$ (this value is re-calculated in each interval $J$ of length $975$)}. In addition, the plots for $\hat{\Omega}_N$ and $\hat{\Omega}_A$ are shown in Fig. \ref{cl1}, considering for the analytical expression the refinement for $q_{n,t}$; observing that for large $\Omega$ the curves become similar, that is, if $t \to -\infty$ the relative error is getting smaller and noticing that for the value $q_{n,t}=1.95203$ defined in the interval $ J=[-4000,-3025]$ when we extend the application to the region $[-6000,-3025]$ the error rate is still acceptable, allow us to consider this quantity, in this interval, constant. 

Finally, Fig. \ref{cl2} and \ref{beta} we present plots of  $\Omega_N$, $\Omega_A$ and $\beta_N$, $-\beta_A$ showing the similarities with the numerical solutions.
\begin{table}[htbp]
\begin{center}
\begin{tabular}{|c|c|c|c|}
\hline
$t$ &  $n$  & $q_{n,t}$ & ER $\Omega$'s  \\
\hline \hline
-200&  $1 $  &   $ 1.80811 $  &   $ 73.7653\% $ \\ \hline
-225&  $1 $  &   $ 1.80811 $  &   $ 57.6173\% $ \\ \hline
-1000&   $ 32$   &  $1.80811$  &  $ 0.0000389\%$ \\ \hline
-1025&   $ 33$   &  $1.90406$  &  $ 5.03481\%$ \\ \hline
-1200&  $ 40$  &   $ 1.90406$   & $ 3.4769\% $\\ \hline
-2000&  $ 72$     &  $  1.90406 $  & $ 0.00038\% $\\ \hline
-2025&  $ 73$     &  $  1.93604 $  & $ 1.61664\% $\\ \hline
-2200&  $ 80$  &  $ 1.93604 $  & $ 1.2169\% $\\ \hline  
-3000&  $ 112 $  &   $  1.93603 $  & $ 0.0001836\% $\\ \hline
-3025&  $ 113 $  &   $  1.95203 $  & $ 0.798575\% $\\ \hline
-3200&  $ 120 $  &   $  1.95203 $  & $ 0.618331\% $\\ \hline
-4000&$ 152$  &   $  1.95203 $  & $ 0.0001362\% $\\ \hline
-5000&$ 192$  &   $  1.95203 $  & $ 0.488977\% $\\ \hline
-6000&$ 232$  &   $  1.95203 $  & $ 0.812403\% $\\ \hline
\end{tabular}
\caption{Error rate between $\hat{\Omega}_N$ and $\hat{\Omega}_A$ where $t_0=-100$, $P_{\beta_{0}}= 2/5$, $\theta=5$, $\eta=0.1$, $C=1.5\times 10^{-5}$ and $\Omega(t_0)=\Omega_0=2.20265\times10^4$, $\beta(t_0)=\beta_{0}=4.68142\times10^4$.}
\label{tab1}
\end{center}
\end{table}

\section{Conclusions and Outlooks} \label{CO}

 In this paper we have presented noncommutative quantum cosmology thorough the help of the WKB-type method for the WDW equation. Here is investigated the homogeneous cosmologies, in a NCC and NCCM frame.
 Although in both models the behavior of the functions (see the integrands in \eqref{int} and \eqref{ks2} in the respective limit) are similar we observe that the possibilities for an asymptotic function could be many (see Appendix \ref{ap.c}) and the selection of the function in the equivalence class vary, being more natural for the FLRW model. The element chosen in the KS metric is relevant in the analysis, since it generates the associated ultrafilter and is not always possible to know it in an explicit way, given the maximality of this family. 

 In the FLRW model, when $\theta\to0$ in $\alpha$ and $\phi$, we obtain the commutative expressions for the general form and 
if $k=0$ in Eqs. \eqref{alpha} and \eqref{phi0} we have the commutative and NCC solutions in the case $k=0, \Lambda \neq 0$ reported previously in literature (see Ref. \cite{1eri}). In addition, making $ \Lambda=0$ together with $\hat{\alpha} \to 3\alpha/2$, $t-t_0 \to 2(\tau-\tau_0)/3$ we return to the results shown previously in Ref. \cite{1eri} for the case $k\neq 0, \Lambda=0$. 

In the KS universe this significance is reflected when the classical system is considered, since in previous works the noncommutativity extension of this model is studied \cite{tachy1,neto,bastos}, extending our analysis and using the analytical form for $\Omega$ we get proposals that fits, in the limit imposed, with the numerical solution of the classical system. In the expression for $\Omega$ we take care to preserve the asymptotical behavior that is reflected in the refinement of the values $q_{n,t}$, but remembering that when $t \to -\infty$ (large $\Omega$) this value can have a constant treatment. Here, is important to take in mind that the presence of the parameter of noncommutativity in momenta could be of relevance for the selection of possible initial states in the early Universe \cite{bastos}. The equation \eqref{r1} is a divergent sum and we use only one significative number to estimate the quotient, this leave the chance to study another expression for $r_n$. For example, the equation:
\begin{equation}
r_n=(a_0\times 10^{-1})+\sum_{2\leq i\leq n}(s_{i,t_i}\times 10^{-i}),
\end{equation}
where $a_0$ is treated like a constant number in the region $R$ (remember that the sets $Z_i$ become to get bigger as $t$ decrease), $t_i \in I_i \subset R$ and $\{s_{i,t_i}\}$ is a convenient increasing sequence.

 When we deal with asymptotic equivalent functions, we have to consider that their relative error is zero as Tables \ref{tab}, \ref{tab1} and Figs. \ref{cl2} ,\ref{beta} shows. Indeed, the comparison between the analytic proposals versus the numerical solutions, the error remains sufficiently small to be able to consider them as a good approximation (the analytical expression and numerical solution can be treated like asymptotically equal). Indeed, we notice that on a complete deformed space it is possible to obtain an expression for the FLRW model, proceeding similarly as in the KS universe. 

Since it have been shown an unexpected connection of some set theoretical concepts with the quantum mechanics as well as in cosmology \cite{set,set1}, the treatment in this scenario via this ideas is the following steps to explore, taking the formal models in ZFC (Zermelo-Fraenkel-Axiom of Choice) and hence the forcing as that special tool to make the shift from the micro to macro scale. Finally, others scenarios in quantum cosmology can be analyzed in order to explore the feasibility of the mathematical methods presented in this paper, however this is work that will be done elsewhere. 

\begin{acknowledgments}
We thank the anonymous referee for thoughtful remarks and suggestions. J. A. A.-M. is supported by CONACyT PhD. grant, also wants to acknowledge the hospitality of UAF-UAZ where part of this work was done. M.A.G.-A. acknowledges support from SNI-M\'exico and CONACyT research fellow. 
Instituto Avanzado de Cosmolog\'ia (IAC) collaborations. We also thank the enlightening conversation with Prof. Salvador Garc\'ia Ferreira.\\
\end{acknowledgments}

\appendix 

\section{Complementary notions}\label{ap.b}
In the following, we extend operationally some concepts treated in Section \ref{intro} to give completeness to our work. First, the noncommutative transformations for the coordinates and their momenta that allow us to convert a noncommutative system into a commutative one depending of the parameters of no commutation\footnote{We have to be careful with the changes that the Moyal product of functions in the minisuperspace $f(x^1,x^2)\star g(x^1,x^2)=f(x^1,x^2)\exp\Big[\frac{i\theta}{2}\Big(\partial_{x^1} \partial_{x^2}-\partial_{x^2} \partial_{x^1}\Big)\Big]g(x^1,x^2)$ represented in the transformations, makes in the quantum equation.} are: 
\begin{equation}
\bar{x}^i=x^i- \frac{1}{2}\theta^{ij}p_{j}, \quad \bar{p}_{j}=p_{j}+ \frac{1}{2}\eta_{ij}x^i, \quad  i,j=1,2, 
\end{equation}
and satisfy the algebra 
\begin{subequations}
\begin{eqnarray}
&\left[\bar{x}^i,\bar{x}^j \right] =i \theta^{ij},\quad \left[\bar{p}_{i},\bar{p}_{j}\right]= i \eta_{ij},\\ &\left[\bar{x}^{i},\bar{p}_{i}\right]=i (\delta_{ij}+\sigma_{ij} ), 
\end{eqnarray}
\end{subequations}
where ${p}_{i}={p}_{x^i}$ and
\begin{equation}
 (\theta^{ij})=
\left( \begin{array}{cc}
0 & \theta  \\
 -\theta &0 \\ 
 \end{array} \right),
 (\eta_{ij})=
\left( \begin{array}{cc}
0 & \eta  \\
 -\eta &0 \\
 \end{array} \right),
 (\sigma_{ij})=\sigma(\delta_{ij}),
  \end{equation}
with $\theta,\eta \in \mathbb{R}$ parameters of noncommutativity in coordinates and momentum, $\sigma=\frac{\theta\eta}{4}$, $(\delta_{ij})$ the identity matrix, and $x^i,p_{i}$ operates in the algebra already known:
\begin{subequations}
\begin{eqnarray}
&\left[x^i,x^j \right] =0,\quad \left[p_{i},p_{j}\right]= 0,\\ 
&\left[x^{i},p_{i}\right]=i.
\end{eqnarray}
\end{subequations}
For the semiclassical scenario the proposed wave function, with the coordinates $x^1,x^2$, is
\begin{equation}
\psi(x^1,x^2)\approx  \exp{i\Big[S_{1}(x^1)+S_{2}(x^2)\Big]},\label{jes0}
\end{equation}
where $S_{1}$, $S_{2}$ takes the dimension of an action for each minisuperspace variable and both satisfy 
\begin{eqnarray}
&\left\vert \frac{d^2 S_i}{d x^{i2}}\right\vert \ll \left(\frac{d S_i}{d x^i}\right)^2, \quad i=1,2.\label{jes}
\end{eqnarray}
To reach the semiclassical limit \cite{lim, lim1} and with \eqref{jes0} we find the following approximations, for $i=1,2$, $k \in \mathbb{R}$ and $n \in \mathbb{N}$
\begin{equation}
\frac{\partial^n \psi}{\partial (x^i)^n}\approx\Big(i \frac{d S_{i}}{d x^i} \Big)^n \psi, \quad e^{k \theta p_{i}}\psi\approx  \Big(1+k \theta\frac{d S_{i}}{d x^i}\Big)\psi, \label{jes1}
\end{equation}
that will be necessary to derive the EHJ equation in the noncommutative context. 

\section{Mathematical background}\label{ap.a}

For our purposes we make use of a particular function and family of subsets described in the following definitions.

Let  $f(x), g(x)$ two functions which are positive when $\lvert x\rvert \to  +\infty$. They are said to be {\it{asymptotically equal}} ($f \sim g$) if
\begin{equation}
\lim_{\lvert x\rvert \to  +\infty}  \frac{ f }{ g  }= 1.\label{asymp1}
\end{equation}
This is an equivalence relation and for $f$ and $g$, belonging to the same class, satisfy the next property derived from the L'H$\rm{\hat{o}}$pital rule.

If $f \sim g$ and $\int_{a}^{ +\infty} g(t)dt=+\infty$ then 
\begin{equation}
\int_{a}^{x} f(t)dt \sim \int_{a}^{x} g(t)dt.  \label{asymp}
\end{equation}
A similar result is obtained when $x \to -\infty$.

Let $X$ a nonempty set with $\lvert X \rvert\geq\omega_0$ and $\mathcal{A}$ a Boolean algebra in $X$. The collection $\mathcal{F}\subset X$ satisfying 
\begin{itemize}
\item{$\emptyset \notin \mathcal{F}$.}
\item{For $A,B \in \mathcal{F}$ we have that $A\cap B\in \mathcal{F}$.}
\item{If $A\in \mathcal{F}$ and $B \in \mathcal{A}$ such that $A\subseteq B$, then $B\in \mathcal{F}$.}
\end{itemize}
is a{\it{ filter}} in $X$. In this work we suppose that  $\mathcal{A}=\mathcal{P}(X)$.  A filter $\mathcal{F} $ is {\it{fixed}}  if $\bigcap \mathcal{F} \neq \emptyset$ and is called {\it{free}} in other case. The filter generated by $A\in  \mathcal{A}$, is $\mathcal{F}_A=\{F \in \mathcal{F} : A\subset F\} $ and a particular case is when $A=\{x\}$ obtaining a {\it{principal filter}}, denoted as $\mathcal{F}_{x}$.\newline
An {\it{Ultrafilter}} in $X$ is a maximal filter $\mathcal{F}$ in the sense that  if we consider any other filter $\mathcal{F}_1$ in X we have that is not finer than $\mathcal{F}$. \newline
A {\it{net}} over $X$ is a map $\varphi:D\to K $ where $D$ is a directed set with a relation $\leq$ and if $\tau$ is a topology in $X$ we say that $\varphi$ converges to $x\in X$ ($\varphi \to x$) if for every $N \in \mathcal{N}_x$, the set of neighborhoods of $x$, there is a $d_N \in D$ with $\varphi(d)\in N$ if $d\geq d_N$.
The following property, related with Ultrafilters, will be of our interest (to check the proof see \cite{com}).\newline
{\bf{Property of finite intersection:}}  Let $C$ a collection of $X$. If for every finite sub-collection $\{A_i : i<\omega_0\} $ we have $\bigcap_{ i<\omega_0}A_i \neq \emptyset$, then there is a fixed Ultrafilter $\mathcal{F}$ with $C\subset \mathcal{F}$.

\section{Asymptotical analysis and Ultrafilters in KS cosmology}\label{ap.c}

In the following lines we present the mathematical development to deal with the problem presented previously. Noticing that one way to extract an adequate asymptotically equal function for the integrand in \eqref{ks2} defined in the interval $(\Omega_0,+\infty)$, $\Omega_0\in \mathbb{R}$, is considering the nonempty infinite set 
\begin{equation}
K=\{f(\Omega) : f\sim G_{\theta,\eta}\},
\end{equation}
therefore, let the collections in $K$
\begin{eqnarray}
A_n=\{ F_{\lambda_{n},c_{1,n},m,c_{2,n}}  : \lambda_{n},c_{1,n},c_{2,n}\in (-n,n),\nonumber\\
m=0,1,\ldots,n\},
\end{eqnarray}
with $n \in \mathbb{N}^+=\mathbb{N}\cup \{0\}, $ $A_0=\{ F_{0,0,0,0} \}$ and $F_{\lambda_{n},c_{1,n},m,c_{2,n}}$ is
\begin{equation}
\Big[\Big(\lambda_{n} P_{\beta_{0}}+\frac{ \eta \Omega}{ 2}  \Big)^2 + c_{1,n}e^{-2m\sqrt{3}\Omega}e^{-\sqrt{3}\theta P_{\beta_{0}} }  +c_{2,n}  \Big]^{-1/2},    
\end{equation}
with $(\lambda_{n},c_{1,n},m,c_{2,n}) \in \mathbb{R}\times \mathbb{R}\times \mathbb{N}^+\times \mathbb{R}$. If  $\mathcal{B}=\bigcup_{n \in \mathbb{N}^+}A_n$ and if we extract a finite sub-collection $\{A_{n_{k}} \}_{k<\omega_0}$ it satisfies the property of finite intersection, then there exists an Ultrafilter in $K$ such that $ \mathcal{B}  \subset \mathcal{F}$. For all $A \in \mathcal{F} \setminus  \mathcal{B} $ is not possible that $A \cap A_0=\emptyset$ showing that $(\forall A \in \mathcal{F} \setminus  \mathcal{B}  \rightarrow F_{0,0,0,0}\in A $) and finally the equality $\mathcal{F}=\mathcal{F}_{ F_{0,0,0,0}}$. Now, if we give the set $K$ the topology $ \tau$ induced by the metric $\rho(f,g)=\sup_{\Omega\in(\Omega_0,+\infty)}\{\lvert (f-g)\rvert \}$ the net $\vartheta:\mathbb{N} \to K$ given by $\vartheta_n=F_{1,1,n,0} \in  \mathcal{B}$ converges (uniformly) to $F_{0,0,0,0}$. Moreover, the map $\varphi:D_{\mathcal{F}_{ F_{0,0,0,0}}}\to K$ defined by $(f,F)\mapsto f$ , where the family
\begin{equation}
\mathcal{D}_{\mathcal{F}_{ F_{0,0,0,0}}}:=\{D=(f,F)  : f\in F\subset \mathcal{F}_{ F_{0,0,0,0}}\},
\end{equation}
is a directed set with the relation $(f_1,F_1)\leq(f_2,F_2)$ iff $F_2\subset F_1$, is a net over $K$ associated to the Ultrafilter $\mathcal{F}_{ F_{0,0,0,0}}$. First, we observe that for every $A \subseteq K$ we have that $A \in \mathcal{F}_{ F_{0,0,0,0}}$ or $X \backslash A \in \mathcal{F}_{ F_{0,0,0,0}}$, denoting $D_A=(f_A,A)$, we have that if $A \in \mathcal{F}_{ F_{0,0,0,0}}$  (in the other case, the treatment is similar) when $D_A \leq D$ implies $f \in A$ and $ \varphi$ is residually in this set (Except, possibly, the constant net $ \varphi(D)=f_0$, for all $D$ in the directed set and $f_0\notin A$.), then for every ball $B(F_{0,0,0,0},\epsilon)$ when 
\begin{equation}
\bigcup\Big(f_\mathcal{C},\mathcal{C}\Big)  \leq D,
\end{equation}
with $\mathcal{C}=B(F_{0,0,0,0},\epsilon)\cap F$ and $F$ in the filter, we get $f \in B(F_{0,0,0,0},\epsilon)$ and
$\varphi \to F_{0,0,0,0}$. In addition, since $K$ is a $T_2$ space, this limit is unique. 
Then, applying the net $\vartheta$ in \eqref{ks2} and making $n\to +\infty$, it is possible to obtain
\begin{align}
\lim_{n\to+\infty}\Big[-\frac{1}{2}\int_ {[\Omega_0,\Omega]} \vartheta_ndx\Big]&=-\frac{1}{2}\int_ {[\Omega_0,\Omega]} \Big( \lim_{n\to+\infty}\vartheta_n\Big)dx\nonumber\\
&=-\frac{1}{2}\int_ {[\Omega_0,\Omega]} F_{0,0,0,0}dx\nonumber\\
&=\frac{1}{\eta}ln\Big(\frac{\Omega_0}{\Omega}\Big)\nonumber\\
&=C-\frac{1}{\eta}ln\Big(\Omega\Big).\label{final}
\end{align}
On the other hand, considering the function $( A^2_\eta \Omega^2 - 144\exp(-\sqrt{3}\theta P_{\beta_{0}}) )^{-1/2}$  where $A_\eta= \sqrt{3} P_{\beta_{0}}+\frac{ \eta }{ 2}$,  $P_{\beta_{0}}\ll 1$ in the final expression, after making an e-folding to $\Omega$ and taking an adequate limit ($\eta \to 0$, or $\eta,\theta \to0$) we recover the noncommutative and commutative expressions already known. We remark, for this part, that the commutative solution solve the Einstein's field equations and the noncommutative one can be derived deforming the symplectic structure at a classical level \cite{neto}, inferring the same in the FLRW model for the mentioned cases. The above leaves the opportunity, applying an appropriate analysis, to find an expression that satisfy another characteristics that could be mathematically or physically relevant.

\begin{figure*}[h]
\centering
\subfigure[$q_{n,t}=1.80811$.]{\includegraphics[width=0.35\textwidth]{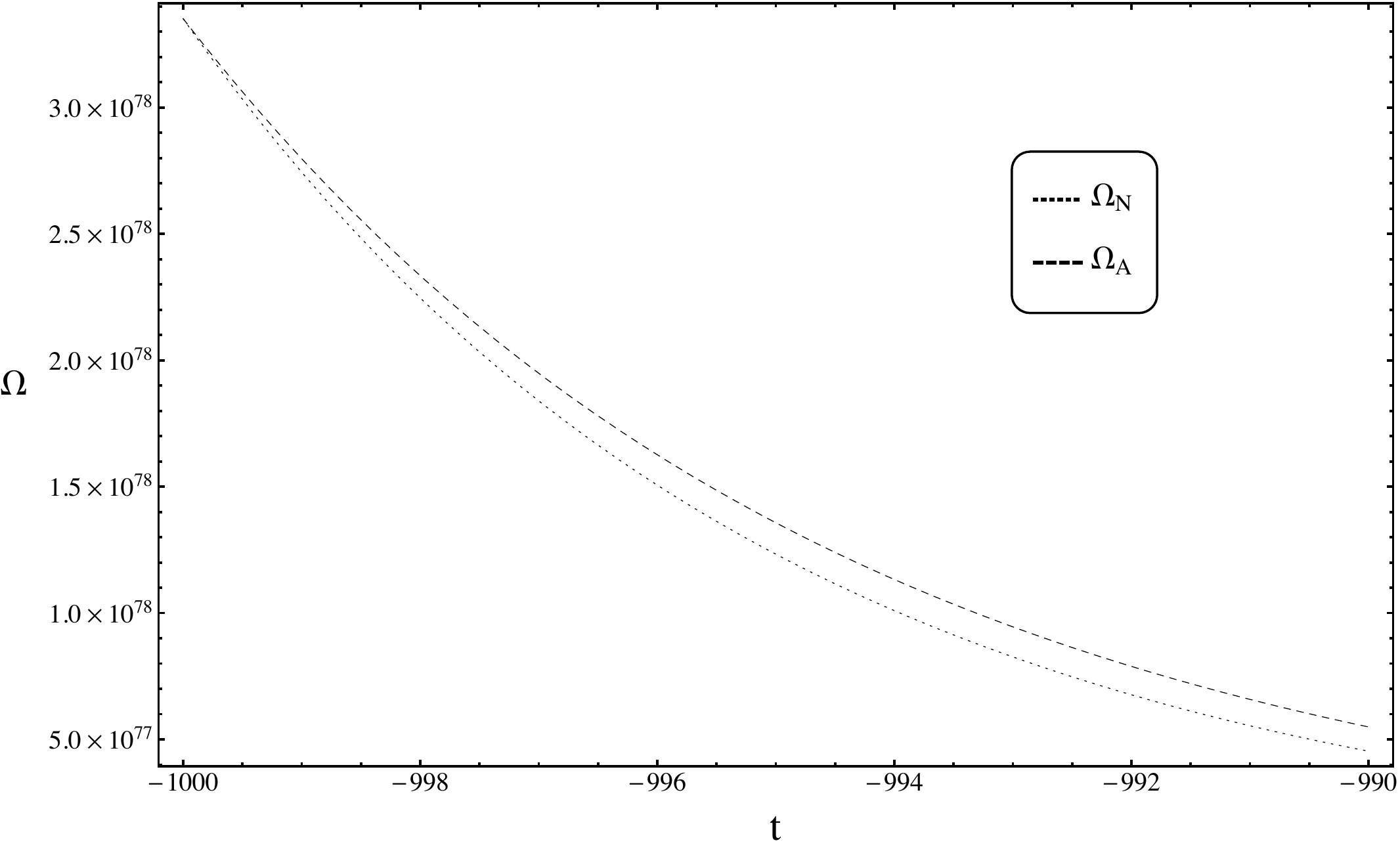}} \label{cla01}
\subfigure[$q_{n,t}=1.90406$.]{\includegraphics[width=0.35\textwidth]{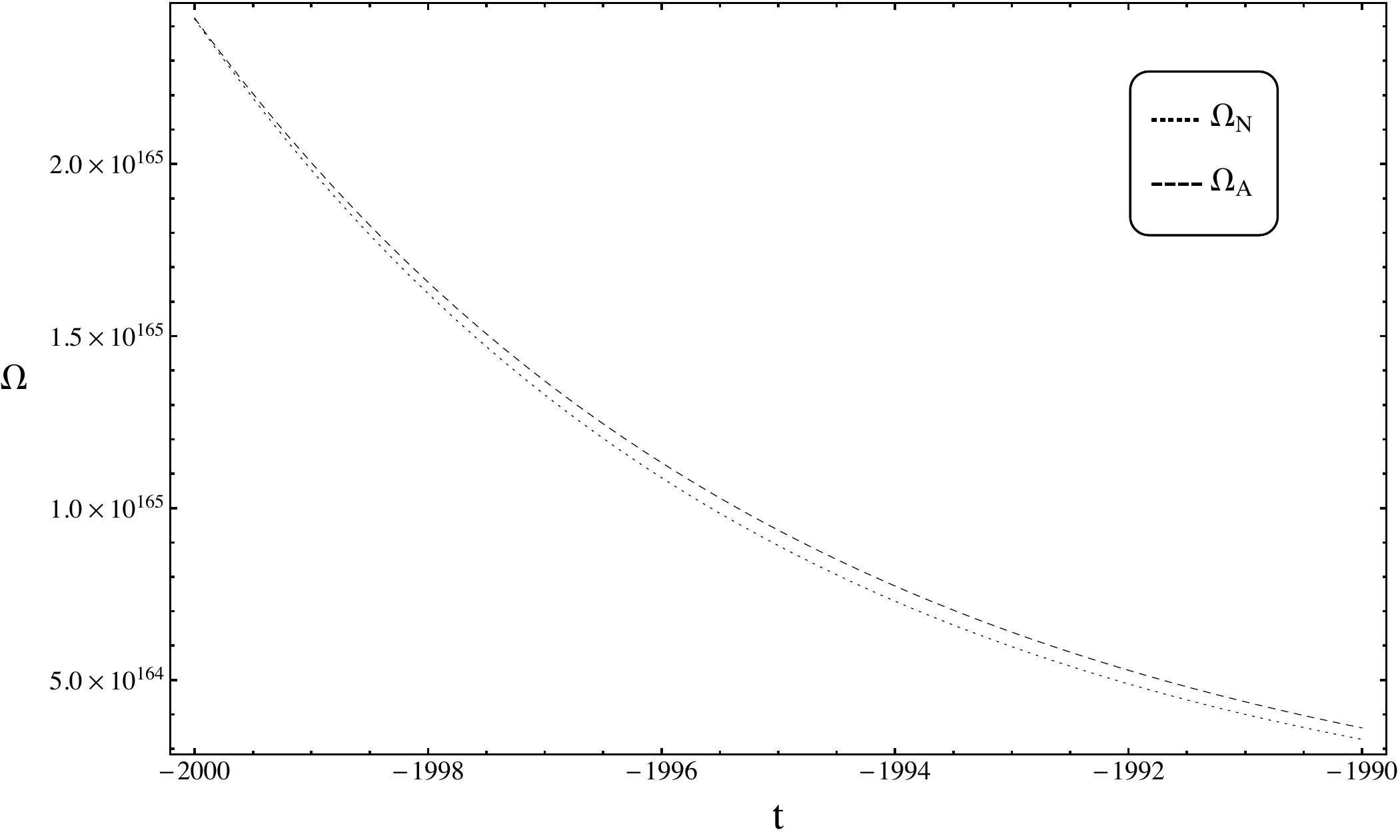}} \label{cla02}
\subfigure[$q_{n,t}=1.93604$.]{\includegraphics[width=0.35\textwidth]{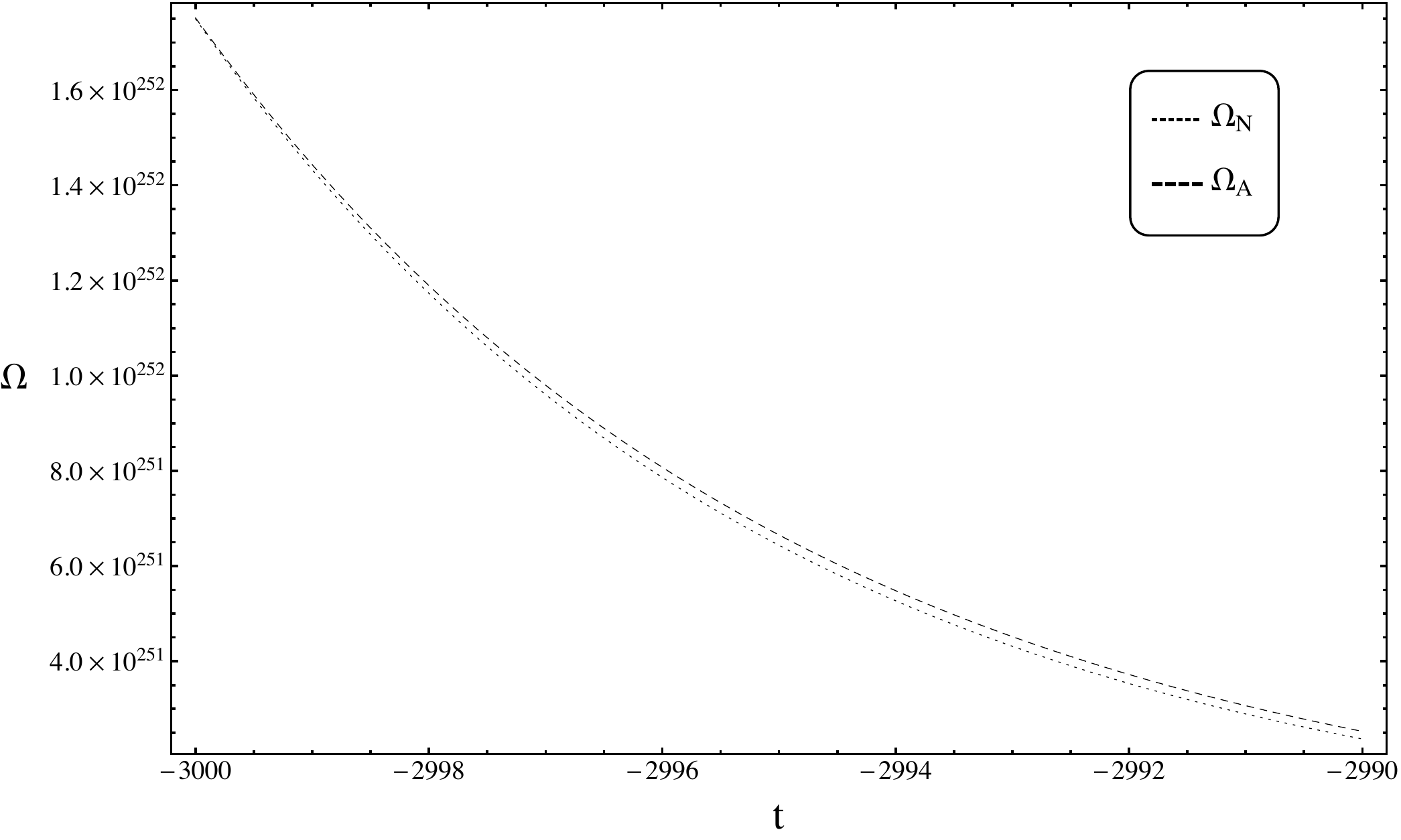}} \label{cla03}
\caption{Plots of $\Omega_N$ (dotted line) and $\Omega_A$ (dashed line) in the intervals $[-1000,-200]$ with $q_{n,t}=1.80811$, $[-2000,-1200]$ with $q_{n,t}=1.90406$ and $[-3000,-2200]$ with $q_{n,t}=1.93604$,
  where $t_0=-100$, $\theta=5$, $\eta=0.1$, $C=1.5\times 10^{-5}$ also considering the initial conditions $\Omega(t_0)=\Omega_0=2.20265\times10^4$ and $\beta(t_0)=\beta_{0}=4.68142\times10^4$, $P_{\Omega}(t_0)= 0$, $P_{\beta}(t_0)= 2/5$. } \label{cl2}
\end{figure*}

\bibliography{librero1}

\end{document}